\documentclass[12pt]{amsart}
\usepackage{amsmath,amsfonts,amssymb}
\usepackage{pdfsync}

\theoremstyle{plain}

\theoremstyle{definition}

\numberwithin{equation}{section}

\def\beq{\begin{equation}}
\def\eeq{\end{equation}}
\def\beqa{\begin{eqnarray}}
\def\eeqa{\end{eqnarray}}

\font\eightrm=cmr8

\def\ga{\gamma}

\def\La{\Lambda}
\def\om{\omega}

\def\sig{\sigma}
\def\ti{\tilde}
\def\vphi{\varphi}

\def\text{\textstyle}

\def\emp{\emptyset}
\def\sub{\subset}
\font\Bbb=msbm10
\let\mcd=\mathchardef
\mcd\Ze="7C5A
\def\O{{\mathcal O}}
\def\12{{\text{\frac{1}{2}}}}
\def\472{{\text{\frac{47}{2}}}}
\def\512{{\text{\frac{51}{2}}}}
\def\352{\text{\frac{35}{2}}}
\def\652{\text{\frac{65}{2}}}
\def\52{\text{\frac{5}{2}}}
\def\72{\text{\frac{7}{2}}}

\newcommand{\half}{\frac{1}{2}}

\begin{document}
\title[Layering in the Ising model]{Layering in the Ising model
}
\author[K.S. Alexander]{Kenneth S. Alexander}
\address{Department of Mathematics KAP 108\\
University of Southern California\\
Los Angeles, CA  90089-2532 USA}
\email{alexandr@usc.edu}
\author[F. Dunlop]{Fran\c cois Dunlop}
\address{Laboratoire de Physique Th{\'e}orique et Modelisation (CNRS,
  UMR 8089)\\ 
Universit{\'e} de Cergy-Pontoise, 95302 Cergy-Pontoise\\
France}
\email{Francois.Dunlop@u-cergy.fr}
\author[S. Miracle-Sol\'e]{Salvador Miracle-Sol\'e}
\address{Centre de Physique Th{\'e}orique, CNRS, Case 907 \\
13288   Marseille cedex 9, France}
\email{miracle@cpt.univ-mrs.fr}


\begin{abstract}
We consider the three-dimensional Ising model in a half-space with a
boundary field (no bulk field). We compute the low-temperature
expansion of layering transition lines.
\end{abstract}
\maketitle

\section{Introduction and results}
We consider the Ising model in the half-space $Z^3_+\sub Z^3$, with
spins $\sig_i=\pm1$, $i\in Z^3_+=\{(i_1,i_2,i_3)$, $i_3\ge1\}$. The
value $-1$ of the spin is associated 
with component or species $A$ of a mixture and the value $+1$ 
is associated with component or species $B$, while the other
half-space $\{i_3\le0\}$ represents a fixed given substrate or wall
$W$, made of a third component or species. The formal Hamiltonian is
\beq\label{HABW}
H^{ABW}=J_{AB}\sum_{<i,j>}(1-\sig_i\sig_j)+J_{WA}\sum_{i_3=1}(1-\sig_i)
+J_{WB}\sum_{i_3=1}(1+\sig_i)
\eeq
with energy contributions $2J_{AB}$, $2J_{WA}$, $2J_{WB}$ associated
respectively to pairs of nearest neighbors $AB$, $WA$, $WB$.
In the first sum, $<i,j>$ are nearest neighbors in $Z^3_+$.
A wetting transition may occur when the bulk phase is $B$ (or
$B$-rich) but the wall prefers $A$: $J_{WA}<J_{WB}$. 

At zero temperature, a macroscopic film of $A$ will separate the wall from the
bulk phase if $J_{WA}+J_{AB}<J_{WB}$. One says that the wall is
``completely wet'' by phase $A$. Raising the temperature will favor
the presence of a film, because the $AB$ interface brings
entropy. Therefore, at positive temperature, a film of $A$ will
always be present if $J_{WA}+J_{AB}\le J_{WB}$. There is no wetting
transition, only complete wetting.  

On the other hand, if $J_{WA}+J_{AB}>J_{WB}$, at zero temperature no
$A$ is present, and at low temperature the wall will be only
partially wet by phase $A$. The density of $B$ tends exponentially
fast to the bulk density of $B$ as a function of the distance to the
wall. Raising the temperature now may produce a
transition from partial to complete wetting: this is the wetting
transition predicted by Cahn \cite{Ca} on the basis of critical
exponents, and then confirmed by numerical and real experiments.

The existence of the wetting transition has been proved mathematically
in the two-dimensional Ising model \cite{A}, but not in the
three-dimensional Ising model. 
Let us simplify the notation to $J=J_{AB}$ and $K=J_{WB}-J_{WA}$, with
\beq
J>0,\qquad 0<K<J.
\eeq
Let $\tau^\pm$ denote the $+/-$ interface tension, defined for the
Ising model in the full space $Z^3$, without wall, with Hamiltonian
equal to the first term of (\ref{HABW}).
Fr\"ohlich and Pfister (see formula (2.20) and Fig. 2 in \cite{FPc})
have proven, among other things:
\beq\label{FP2}
K<\frac{1}{2}\tau^\pm\qquad\Longrightarrow\qquad{\rm Partial\ wetting}.
\eeq
This is a non-perturbative result, valid for all temperatures $0\le T<T_c$.

We shall consider only low temperatures, and perturbative arguments
(not fully mathematically rigorous),
indicating that the partial wetting range is slightly wider than (\ref{FP2}),
and includes first order layering transitions, as we now explain.
Consider the model in a box $\La\sub Z^3_+$, with bottom layer at
$i_3=1$, and boundary condition $\bar\sig$ on the other
five sides of the box. Let $\La_1=\La\cap\{i_3=1\}$. 
The Hamiltonian (\ref{HABW}) may be cast into the equivalent form
\beq
H_\La(\sig_\La|\bar\sig)=-2J|\La_1|+J\sum_{<i,j>\cap\La\ne\emp}(1-\sig_i\sig_j)
+K\sum_{i_3=1}(1+\sig_i).
\eeq
In the first sum, $i,j$ are nearest neighbors in $Z^3_+$
(so neither $i$ nor $j$ is in the wall), and $\sig_i$
or $\sig_j$ should be replaced by $\bar\sig_i$ or $\bar\sig_j$ wherever
$i\not\in\La$ or $j\not\in\La$. In the second sum, $i\in\La$. The
constant term in front is a convenient normalization.
Boundary condition $n$, with $n=0,1,2,\dots$, is associated with the
configuration $n$ in $Z^3_+$, given by
\beq\label{barsig}
\bar\sig_i=-1\ {\rm if}\ i_3\le n,\qquad\bar\sig_i=+1\ {\rm if}\ i_3>n.
\eeq
A possible scenario for the wetting transition is as follows (see Fig. 1):
Let $0<K<J$ with $J-K$ small. At $T=0$ we have configuration 0, and
for small $T$, we are close to configuration 0, call it state 0: in
the thermodynamic limit, the probability that at a given $i$ the spin
$\sig_i$ differs from $\bar\sig_i$, defined by (\ref{barsig}) with
$n=0$, is small. State $n$ is defined similarly from
configuration $n$, for any $n$. As the temperature is raised, a first 
order transition will occur, from state 0 to state 1, then as the
temperature is raised further, from state 1 to state 2, and so on. The
level of the stable state $n$ goes to infinity as the temperature
approaches the wetting transition temperature, which in this case is
strictly below the roughening temperature.  
This scenario, with a sequence of first order layering transitions
leading to the wetting transition, is part of the general picture
which emerged based upon various physical heuristics and Monte-carlo
simulations (see \cite{BL,PSW} and references therein.)
\font\eightrm=cmr8
\begin{center}
\setlength{\unitlength}{.7mm}
\begin{picture}(0,70)

\put(-50,0){\vector(0,1){60}}
\bezier{30}(30,0)(30,20)(30,40)
\put(-50,0){\vector(1,0){130}}
\put(-60,55){$T$}
\put(74,2){$K$}
\put(40,-2){\line(0,1){2}}
\put(41,2){$J$}
\bezier{200}(-20,46)(40,13)(40,0)
\bezier{200}(-3,41)(40,13)(40,0)
\bezier{200}(10,37)(40,13)(40,0)
\bezier{200}(18,34)(40,13)(40,0)
\bezier{200}(23,32)(40,13)(40,0)
\bezier{200}(26,31)(40,15)(40,0)
\put(34,32){{\eightrm complete}}
\put(37,27){{\eightrm wetting}}
\put(0,15){\eightrm 0}
\put(-7,39){\eightrm 1}
\put(8,34){\eightrm 2}
\put(17,31){\eightrm 3}
\end{picture}
\end{center}
\bigskip
\begin{center}
\begin{minipage}{9cm}
{\footnotesize
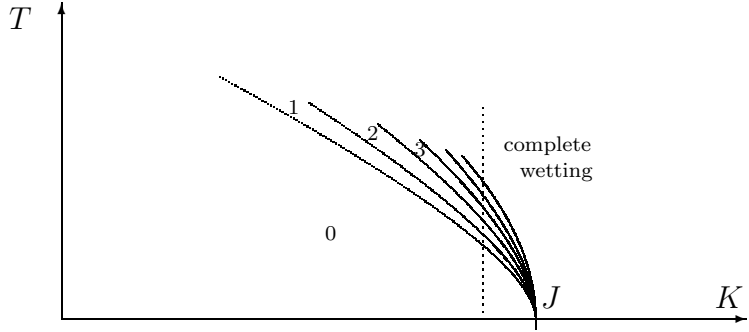
Fig. 1. Layering transition lines near $T=0$. Dotted line shows a path
from partial to complete wetting.}
\end{minipage}
\end{center}

\bigskip
Let $t=e^{-4\beta J}\ll1$ and $u=2\beta(J-K)=\O(t^2)$.  Note that
each factor of $t$ corresponds to two plaquettes of the interface.
We find the following approximation to the coexistence (first order
transition) lines starting from $(t=0,u=0)$: 
\begin{align}\label{123456}
&0/1:\quad u=-\ln(1-t^2)+t^3+\O(t^4)\notag \\
&1/2:\quad u=-\ln(1-t^2)-t^3+5t^4+\O(t^5)\notag \\
&2/3:\quad u=-\ln(1-t^2)-t^3+4t^4-4t^5+\O(t^6)\notag \\
&3/4:\quad u=-\ln(1-t^2)-t^3+4t^4-6t^5+{\text\frac{51}{2}} t^6+\O(t^7)\notag \\
&4/5:\quad u=-\ln(1-t^2)-t^3+4t^4-6t^5+\472 t^6-51t^7+\O(t^8)\notag \\
&5/6:\quad u=-\ln(1-t^2)-t^3+4t^4-6t^5+\472 t^6-53t^7+162t^8+\O(t^9)\notag \\
&6/7:\quad u=-\ln(1-t^2)-t^3+4t^4-6t^5+\472 t^6-53t^7+160t^8\notag \\
&\qquad \qquad \qquad + (B_9+2)t^9 + \O(t^{10})\notag \\
&7/8:\quad u=-\ln(1-t^2)-t^3+4t^4-6t^5+\472 t^6-53t^7+160t^8\notag \\
&\qquad \qquad \qquad + B_9t^9 + \O(t^{10})
\end{align}
Here $B_9$ is a constant which we do not calculate, but we show it is the same for all 
interface heights $n \geq 6$.  The analogous statement applies to the calculated coefficients
as well, for example, the coefficient of $t^4$ is 4 for all $n \geq 2$.
This is a result of the cancellation of all terms proportional to $n, n^2$, etc. in
the low temperature expansion of the increment of surface free energy
from $n$ to $n+1$, up to the given orders in $t$.
We are unable to determine a systematic 
way in which this cancellation occurs, but we anticipate its validity for all orders in $t$.
The consequence is that each successive transition line requires one more order in $t$
to discern it.

The phases 0, 1, 2, 3, 4, 5, 6, 7 are predicted to be stable between the
respective transition lines. In particular phase 0 should be stable for
$u>t^2+t^3+\O(t^4)$. For comparison, (\ref{FP2}) gives partial wetting for 
$u>2t^2+4t^3+\O(t^4)$.
Basuev \cite{Ba} has given such equations for coexistence of the
phases 0,1,2 with 1,2,3 respectively.  

Naturally, more is known in the SOS approximation, and in that context
full mathematical rigor is possible, see \cite{Ch,ADM}.
The low-temperature expansions of the Ising model and the
corresponding SOS model agree only up to and including order $t^2$,
which is of little help for (\ref{123456}). 
Order $t^3$ corresponds to a domino excitation of the interface, same
in Ising and SOS, but also to a unit cube bubble, present only in the
Ising model.

The stability range of phase $n$ appears to be of width approximately
$2t^{n+2}$ in the variable $u$. This is the same for Ising and SOS, and is the
result of a double leg interface excitation reaching the wall (see Fig. 7).

The $n/n+1$ coexistence lines are expected to converge as $n\to\infty$
to a part of the wetting transition line. Therefore the low-temperature
expansion of the $n/n+1$ coexistence lines for all $n$ would give the
low-temperature expansion of the  wetting transition line. 

The derivation of the 2/3, 3/4, 4/5, 5/6, 6/7 transition lines is given in
Section \ref{LTE}, except for the recursion diagrams, which are
displayed and explained in Section \ref{RecDia}. The special features
of the 0/1 and 1/2 transition lines are given in Section \ref{012}.
Diagrams for the 7/8 transition line are postponed to Section \ref{78}.

\section{Low temperature expansion}\label{LTE}
Let us consider a finite volume and boundary condition $n$, with
$n\ge1$ for definiteness. The ground state is (\ref{barsig}),
with a flat interface at height $n+{\frac{1}{2}}$, denoted $I_n$. 
At positive temperature, bubbles and interface excitations will
appear. If state $n$ is stable, or if the statistical ensemble is restricted
by a condition forbidding large fluctuations, the gas of bubbles and
interface excitations should be diluted, and the corresponding dilute gas
expansion is expected to give exact asymptotics for low temperatures. 
The corresponding partition function is
\beq\label{Zndef}
Z_n^\Lambda = {\sum_{\sigma_\Lambda}}' e^{-\beta H_\La(\sig_\La|n)},
\eeq
where $\beta = 1/kT$ is the inverse temperature and the $\,'\,$ indicates
that summation is over a restricted ensemble corresponding to state $n$.
The associated surface free energy density (times $\beta$) will be
denoted $f_n$, so that
\beq
f_n-f_{n+1}=
\lim_{\La\nearrow Z_+^3}-\frac{1}{|\La_1|}\log\frac{Z_n^\La}{Z_{n+1}^\La}
\eeq
We are going to compute the leading terms up to some order for 
$f_n-f_{n+1}$, so as to obtain (\ref{123456}). 

Bubbles and interface excitations will be called contours, or also
polymers, and will be denoted $\ga$. They are defined 
as boundaries of maximal connected sets of points where the spin
differs from its ground state value in the corresponding restricted
ensemble. A set of points is connected 
if any two points can be connected by a path of nearest neighbor
bonds in the set. The boundary of a set of points is a set of plaquettes.
A contour need not be connected.
Interface excitations are distinguished by the property of sharing at
least one plaquette with $I_n$. A bubble crossing $I_n$ without
sharing a plaquette is not an interface excitation.

The low-temperture polymer expansion starts with
\beq\label{Zn}
Z_n^\La=e^{u|\La_1|\delta(n)}\sum_{\{\ga\}}\prod_\ga \vphi(\ga)
\eeq
where $\{\ga\}$ is a compatible family of contours, and $\vphi(\ga)$ is
the weight of a contour,
\beq\label{phi}
\vphi(\ga)=t^{{\frac{1}{2}}|\ga|-|\ga\cap I_n|}
e^{u|\ga\cap\{z={\frac{1}{2}}\}|}
\eeq
where $|\cdot|$ is the number of plaquettes in $\ga$ or in $\ga\cap
I_n$ or in $\ga\cap\{z={\frac{1}{2}}\}$.
A family is compatible if any pair of contours in the family is
compatible. Two contours are compatible if their interiors are
disjoint and they share no plaquette.
In view of (\ref{phi}), we will represent an interface excitation with
plaquettes in $I_n$ removed (see Fig. 2-7 below), but when deciding
compatibility, it must be remembered that these plaquettes
do belong to the interface excitation.

As the interaction between contours is a two-body interaction ---
compatibility is decided two by two --- the general theory of polymer
expansion (see e.g. \cite{GMM,KP,M}) gives, from (\ref{Zn}), 
\beq\label{logZn}
\log(Z_n^\La)=\sum_\om\vphi^T(\om)
\eeq
where $\om$ is a cluster or family of contours,
with contour $\ga$ repeated $n_\ga$ times, and
\beq\label{phiT}
\vphi^T(\om)=\prod_{\ga\in\om}\Bigl(\frac{1}{n_\ga!}\vphi(\ga)^{n_\ga}\Bigr)
\sum_G(-1)^l
\eeq
where the sum over $G$ is over connected graphs
on the cluster, and $l$ is the number of edges in $G$. An edge
may exist between $\ga$ and $\ga'$ if and only if $\ga$ and $\ga'$ are
incompatible.

For the expansion of $\tau^\pm$,
interface excitations were expanded in terms of walls and ceilings by 
Dobrushin \cite{Do}, who proved convergence of the resulting expansion. For the
SOS approximation of the present wetting model,  
a two-scale convergent expansion was used in \cite{ADM}.
Here we consider only the finite volume expansion and the formal
infinite volume series, which is why our derivation of \eqref{123456}
is not fully rigorous. 

All the clusters in (\ref{logZn}) 
lie within $\La$.
For a  cluster which contains an interface excitation, we write
$\om\in I_n$. For a cluster of bubbles only, compatible (i.e. not sharing a plaquette)
with $I_n$, we write $\om\sim I_n$.
For a  cluster which reaches the bottom $\{i_3=1/2\}$, 
we write $\om\in W$, otherwise $\om \approx W$.  We write $W_N$ for 
the top boundary $\{i_3 = N+\half\}$ of $\Lambda$.
All clusters $\om\subset\Lambda$ are compatible with
the top boundary; we write $\om\approx W_N$. Then
\beqa\label{4term}
\log(Z_n^\La)&=&\sum_{\om\in I_n,W\atop\om\approx W_N}\vphi^T(\om)+
\sum_{\om\in I_n,\atop\om\approx W, W_N}\vphi^T(\om)+
\sum_{\om\in W,\atop\om\sim I_n,\om\approx W_N}\vphi^T(\om)+
\sum_{\om \approx W,W_N\atop \om\sim I_n}\vphi^T(\om)\cr
&=&\sum_{\om\in I_n,W\atop\om\approx W_N}\vphi^T(\om)+
\sum_{\om\in I_n,\atop\om\approx W,W_N}\vphi_0^T(\om)+
\sum_{\om\in W,\atop\om\sim I_n,\om\approx W_N}\vphi_1^T(\om)+
\sum_{\om \approx W,W_N\atop \om\sim I_n}\vphi_2^T(\om)
\eeqa
where
\beq\label{phi012}
\vphi_0(\ga)=t^{{\frac{1}{2}}|\ga|-|\ga\cap I_n|}\,,\qquad
\vphi_1(\ga)=t^{{\frac{1}{2}}|\ga|}e^{u|\ga\cap\{z={\frac{1}{2}}\}|}\,,\qquad
\vphi_2(\ga)=t^{{\frac{1}{2}}|\ga|}.
\eeq
The first term in (\ref{4term}) depends explicitly upon $n$.
The sums consist of clusters $\om\subset\Lambda$, but
in order to extract the $n$-dependent part of the following three terms, 
it is convenient
to relax this condition into $\om\cap\La\ne\emp$,
allowing ``boundary-overlapping'' clusters which overlap $W$ or $W_N$. 
In this context the notations $\om \approx W, \om \in W$ and $W \approx W_N$
apply only to clusters 
which do not overlap $W$ and $W_N$ respectively.  
Then applying inclusion-exclusion to the summation conditions, the last three sums
in \eqref{4term} become
\begin{align}\label{3term}
\sum_{\om\in I_n,\atop\om\approx W, W_N}&\vphi^T(\om) =
  \sum_{\om\in I_n}\vphi_0^T(\om)-
  \sum_{\om\in I_n,\atop\om\not\approx W}\vphi_0^T(\om)-
  \sum_{\om\in I_n,\atop\om\not\approx W_N}\vphi_0^T(\om)+
  \sum_{\om\in I_n,\atop\om\not\approx W, W_N}\vphi_0^T(\om)\notag \\
\sum_{\om\in W,\atop\om\sim I_n,\om\approx W_N}&\vphi^T(\om) =
  \sum_{\om\in W}\vphi_1^T(\om)-
  \sum_{\om\in W,\atop\om\not\sim I_n}\vphi_1^T(\om)-
  \sum_{\om\in W,\atop\om\not\approx W_N}\vphi_1^T(\om)+
  \sum_{\om\in W,\atop\om\not\sim I_n,\om \not\approx W_N}\vphi_1^T(\om)\notag \\
\sum_{\om\approx W,W_N,\atop \om \sim I_n}&\vphi^T(\om) =
  \sum_{\om\cap\La\ne\emp}\vphi_2^T(\om)-
  \sum_{\om\not\sim I_n}\vphi_2^T(\om)-
  \sum_{\om\not\approx W}\vphi_2^T(\om)-
  \sum_{\om\not\approx W_N}\vphi_2^T(\om)\notag \\
&+\sum_{\om\not\sim I_n,\atop \om \not\approx W}\vphi_2^T(\om)
  +\sum_{\om\not\sim I_n,\atop W \not\approx W_N}\vphi_2^T(\om)
  +\sum_{\om\not\approx W,W_N}\vphi_2^T(\om)
  -\sum_{\om\not\approx W,W_N\atop \om\not\sim I_n}\vphi_2^T(\om).
\end{align}
Note that the sums from \eqref{4term}, on the left side in \eqref{3term}, are not affected by 
the relaxation from $\om\subset\La$ to $\om\cap\La\ne\emp$.
Terms with $\om\not\sim I_n,\om\not\approx W_N$ or $\om\not\approx W, W_N$ or 
$\om\not\approx W,W_N, \om\not\sim I_n$ are negligible 
in the thermodynamic limit and will be omitted in the sequel.
This is the meaning of $\simeq$ instead of = below. Apart from these
negligible terms, only one sum on the right side each of the three equalities in
(\ref{3term}) actually depends upon $n$. Therefore 
\beq \label{ndepend}
\log(Z_n^\La)\simeq\sum_{\om\in I_n,W}\vphi^T(\om)
-\sum_{\om\in I_n,\atop\om\not\approx W}\vphi_0^T(\om)
-\sum_{\om\in W,\atop\om\not\sim I_n}\vphi_1^T(\om)+
\sum_{\om\not\sim I_n,\atop \om\not\approx W}\vphi_2^T(\om)+{\rm indep.\ of\ }n.
\eeq
In order to compare $Z_n^\La$ and $Z_{n+1}^\La$ using translation invariance, 
the wall $W$ will be denoted $W_0$, and $W_{-1}$ will denote a wall
translated vertically by $-1$.
The following is immediate from \eqref{ndepend}.

\bigskip\noindent{\bf Proposition 1:} For $n\ge1$, in the limit of a
box $\La$ of height $N\to\infty$,
\beqa\label{5term}
\log(Z_n^\La/Z_{n+1}^\La)=\sum_{\om\in I_n,W}\vphi^T(\om)
-\sum_{\om\in I_n,\atop{\om\not\approx W_0\atop\om\approx W_{-1}}}\vphi_0^T(\om)
-\sum_{\om\in I_{n+1},W}\vphi^T(\om)\hskip2cm\cr
-\Bigl(\sum_{\om\in W,\atop{\om\not\sim I_n\atop\om\sim I_{n+1}}}\vphi_1^T(\om)
-\sum_{\om\in W,\atop{\om\not\sim I_{n+1}\atop\om\sim I_n}}\vphi_1^T(\om)\Bigr)
+\Bigl(\sum_{\om\not\approx W,\atop {\om\not\sim I_n}}\vphi_2^T(\om)
-\sum_{\om\not\approx W,\atop{ \om\not\sim I_{n+1}}} \vphi_2^T(\om)\Bigr).\cr
\eeqa
\smallskip
In terms of surface free energy densities, anticipating a leading term
$t^{2n}$, this can be written as
\beqa\label{AB}
t^{-2n}(f_{n+1}-f_n)
&=&A_n(u)-A_n(0)-t^2A_{n+1}(u)-B_n(u)+B_n^\infty(0)
\eeqa
where each of the five terms is defined by the corresponding term in
(\ref{5term}). We can simplify $B_n^\infty(0)$ as follows.  The terms in 
$B_n^\infty(0)$ correspond to 
clusters of bubbles only, and the set of such clusters may be divided
into equivalence classes consisting of clusters which are vertical translates of one another.
Within each equivalence class there is a unique special bubble $\om$ satisfying 
$\om \in W$.  For a given equivalence class, the number of terms from that class
in the first sum
in $B_n^\infty(0)$ is the number of heights $k \geq n$ for which the special bubble has
a horizontal plaquette at height $k+\frac{1}{2}$, and similarly for the second sum,
but with heights $k \geq n+1$.
Hence the net number of terms in $B_n^\infty(0)$ from the equivalence class, counted with $+/-$ sign,
is 1 if the special bubble $\om$
has a horizontal plaquette at height $n+\frac{1}{2}$
(that is, if $\om \not\sim I_n$), and 0 otherwise.
It follows that 
\beqa\label{Bninfty}
  t^{2n}B_n^\infty(0) = \sum_{\om\in W\atop\om\not\sim I_n}\vphi_2^T(\om)
    =t^{2n}\sum_{m=0}^\infty t^{2m}B_{n+m}(0).
\eeqa
Since 
\[
  t^{2n}B_n(u) =\sum_{\om\in W,\atop{\om\not\sim I_n}}\vphi_1^T(\om)
    -\sum_{\om\in W,\atop\om\not\sim I_{n+1}}\vphi_1^T(\om),
\]
and since $\vphi_1^T=\vphi_2^T$ for $u=0$, we have $B_n^\infty(0)=B_n(0)+t^2B_{n+1}^\infty(0)$ so that
\begin{align}\label{AB1}
t^{-2n}&(f_{n+1}-f_n) \notag \\
&=A_n(u)-A_n(0)-t^2A_{n+1}(u)-\bigl(B_n(u)-B_n(0)-t^2B_{n+1}^\infty(0)\bigr).
\end{align}

The $u$ dependence may be written as
\beqa\label{APQRS}
A_n(u)&=&e^uP_n+e^{2u}Q_n+e^{3u}R_n+e^{4u}S_n+e^{5u}T_n+e^{6u}U_n+\dots\\
B_n(u)&=&e^u\ti P_n+e^{2u}\ti Q_n+e^{3u}\ti R_n+e^{4u}\ti S_n+e^{5u}\ti T_n+e^{6u}\ti U_n+\dots
\label{BPQRS}
\eeqa
For $n\ge3$ we have $P_n=\O(1)$ corresponding to interface fluctuations placing 
a single plaquette on the wall, and similarly
$Q_n=\O(t^2)$, $R_n=\O(t^4)$,
$S_n=\O(t^5)$, $T_n=\O(t^7)$, $U_n=\O(t^8)$.
Relative to these,
$\ti P_n,\ti Q_n,\ti R_n,\ti S_n$ have an extra factor $t$ at leading order.  
The remainder in (\ref{APQRS}), (\ref{BPQRS}) is $\O(t^{10})$.
For $n=2$ we have $P_2=\O(1)$, $Q_2=\O(t^2)$, $R_2=\O(t^4)$,
$S_2=\O(t^4)$, $T_2=\O(t^6)$, $U_2=\O(t^6)$,
while $\ti P_2,\ti Q_2,\ti R_2,\ti S_2$ are of the same order as for $n\ge3$. 
The remainder in (\ref{APQRS}) for $n=2$ is $\O(t^8)$,
but in (\ref{BPQRS}) it is still $\O(t^{10})$.

Let $Q_n=Q_n^1+Q_n^2$ and $R_n=R_n^1+R_n^2+R_n^3$, 
where the upper index $1,2,3$ is the number of polymers (in the cluster) 
touching the wall, so that $Q_n^1=\O(t^2)$, $Q_n^2=\O(t^3)$, etc. 
We are going to expand (\ref{AB}) up to order $t^9$, requiring
$A_{n+1}$ up to order $t^7$, using recursion in $n$.

\bigskip\noindent{\bf Recursion:} For $n\ge2$,
\begin{align}\label{induc}
P_{n+1}&=P_n+2Q_n+3R_n+4S_n+5T_n+6U_n-t\bigl(P_n+2Q_n+3R_n+4S_n\bigr)\\
&\quad+\O(t^7)\notag\\
Q_{n+1}^1&=(4t^2-4t^3)P_n+(t+6t^2-7t^3)Q_n^1+(8t^2-8t^3)Q_n^2\notag\\
&\quad+2tR_n^1+9t^2R_n+tR_n^2+4tS_n+\O(t^7)\notag\\
Q_{n+1}^2&=(-5t^3+5t^4)P_n+(-10t^3+10t^4)Q_n^1+t^2Q_n^2-12t^3Q_n^2
+\O(t^7)\notag\\
Q_{n+1}&=(4t^2-9t^3+5t^4)P_n+(t+6t^2-17t^3+10t^4)Q_n^1+(9t^2-20t^3)Q_n^2\notag\\
&\quad+2tR_n^1+9t^2R_n+tR_n^2+4tS_n+\O(t^7)\notag\\
R_{n+1}^1&=(18t^4-18t^5)P_n+(6t^3+24t^4)Q_n^1+t^2R_n+4t^2S_n+\O(t^7)\notag\\
R_{n+1}^2&=(-48t^5+48t^6)P_n-8t^4Q_n^1+\O(t^7)\notag\\
R_{n+1}^3&=31t^6P_n+\O(t^7)\notag\\
R_{n+1}&=(18t^4-66t^5+79t^6)P_n+(6t^3+16t^4)Q_n^1+t^2R_n+4t^2S_n+\O(t^7)\notag\\
S_{n+1}&=(4t^5+60t^6)P_n+2t^4Q_n+\O(t^7), \notag
\end{align}
and the same recursion relations for $\ti P_n,\ti Q_n,\ti R_n,\ti S_n$,
with an error $\O(t^8)$.
The recursion relations (\ref{induc}) have been found with the help of
diagrams, see Section \ref{RecDia}.

Solving these recursion relations for formal power series in $t$ requires 
as input $P_n$, or $\ti P_n$, for all $n$, to the required
order. Indeed the order obtained in the output $P_{n+1}$ or $\ti
P_{n+1}$ is the same as in the input $P_n$ or $\ti P_n$, so that the
recursion formula does not help. On the 
other hand, if the power series expansion for $P_n$ or $\ti P_n$ is
obtained by other methods, up to the required order, for all $n$, then the
initial condition, at $n=2$, given by
$Q_2^1=4t^2+2t^2+\O(t^3)$, $Q_2^2=-5t^3+\O(t^4)$,
$R_2^1=\O(t^4)$, $R_2^2=\O(t^5)$, $S_2=\O(t^4)$, or
$\ti Q_2^1=4t^3+\O(t^4)$, $\ti Q_2^2=-5t^4+\O(t^5)$, $\ti R_2^1=18t^5+\O(t^6)$,
together with $P_n$ or $\ti P_n$, will give one more order in $t$ with
each recursion step. The recursion equation 
giving $P_{n+1}$ or $\ti P_{n+1}$ may be  checked at the end for consistency.  
The final result of this for $P_n,Q_n,R_n,S_n$ or 
$\ti P_n,\ti Q_n,\ti R_n,\ti S_n$ is given as ``first excitations'':

\bigskip\noindent{\bf First excitations in $A_n(u)$:}
\begin{align}\label{PQRS}
P_n&=1-(n-5)t+c_nt^2-a_nt^3+d_nt^4+\O(t^5)\,,\qquad n\ge5\\
Q_n^1&=4\bigl[t^2-(n-6)t^3+(c_{n-1}+7)t^4
-(a_{n-1}+c_{n-1}-c_{n-2}+6n-41)t^5\notag\\
&+(d_{n-1}+a_{n-1}-a_{n-2}+5c_{n-2}+c_{n-3}+2n+C)t^6\bigr]
+2t^n+\O(t^7)\,,\, n\ge6\notag\\
Q_n^2&=-5t^3+5(n-5)t^4-(5c_n+C)t^5+\O(t^6)\,,\qquad n\ge3\notag\\
R_n^1&=18t^4-(18n-114)t^5+(18c_n - 24n+C)t^6+\O(t^7)\,,\qquad n\ge4\notag\\
R_n^2&=-48t^5+(48n+C)t^6+\O(t^7)\,,\qquad n\ge3\notag\\
S_n&=4t^5-(4n+C)t^6+\O(t^7)\,,\qquad n\ge3\notag\\
T_n&=Ct^7 + \O(t^8)\,,\qquad n\ge3,\notag
\end{align}
with
\beqa\label{cn}
c_n&=&{{n-1} \choose 2}+4(n-2)+16,\\ \label{an}
a_n&=&{{n-1} \choose 3}+12{{n-1} \choose 2}-10n-48,\\ \label{dn}
d_n&=&{{n-1}\choose 4} + 20{{n-1}\choose 3} + 32{{n-1}\choose 2} + 54n + C.
\eeqa
In \eqref{PQRS} and in what follows, $C$ is a generic constant, not depending on
$n$ and different at different appearances, which we do not calculate or use.
The expansion for $P_n$ is valid for $n=3$ with two orders less
(that is, $\O(t^3)$ instead of $\O(t^5)$), and
for $n=2$ with three orders less, and for $n=1$ with four orders less.
It is obtained by listing diagrams---see below.
The results for $Q_n^1,\dots S_n$ in \eqref{PQRS} follow from the
result for $P_n$ using (\ref{induc}).
One can start the induction from $n=1$ with (\ref{induc}) adjusted for $n=1$,
or from $n=2$ with $Q_2^1=4t^2+2t^2+\O(t^3)$, $Q_2^2=-5t^3+\O(t^4)$,
$R_2^1=\O(t^4)$, $R_2^2=\O(t^5)$, $S_2=\O(t^4)$.
The expansion for $Q_n^1$ is valid for $n=4$ with one order less, and
for $n=3$ with two orders less, and for $n=2$ with three orders less. 
The expansion for $R_n^1$ is valid for $n=3$ with one order less.

The result for $P_n$ is displayed in Figs 2-6.
Formula (\ref{cn}) for $c_n$ was obtained using Fig. 2 and Fig. 3 with
(\ref{phiT}). The factor 6 for the last diagram in Fig. 3 is: one
incompatible unit cube upward interface excitation, as drawn, and five
incompatible unit cube downward interface excitations.
Formula (\ref{an}) for $a_n$ was obtained using Fig. 4, 5, 6 with (\ref{phiT}).
The factor $5(n-2$) for the one before last diagram in Fig. 4 is: one
incompatible unit cube bubble in the interface leg at height
$2,\dots,n-1$, as drawn, and four incompatible unit cube  bubbles
adjacent to the leg at height $2,\dots,n-1$, and similarly for
the last diagram in Fig. 4.
Formula (\ref{dn}) was obtained using the diagrams in Fig.~13 and Fig.~14.

\vskip0.8cm
\setlength{\unitlength}{2000sp}%
\begin{picture}(12174,2274)(-11,-1423)
\thinlines
{\put(  1,839){\line( 1, 0){450}}
\put(451,839){\line( 0,-1){2205}}
\put(451,-1366){\line( 1, 0){450}}
\put(901,-1366){\line( 0, 1){2205}}
\put(901,839){\line( 1, 0){450}}
}%
{\put(8326,839){\line( 1, 0){450}}
\put(8776,839){\line( 0,-1){1350}}
\put(8776,-511){\line( 1, 0){450}}
\put(9226,-511){\line( 0,-1){855}}
\put(9226,-1366){\line( 1, 0){450}}
\put(9676,-1366){\line( 0, 1){1305}}
\put(9676,-61){\line(-1, 0){450}}
\put(9226,-61){\line( 0, 1){900}}
\put(9226,839){\line( 1, 0){900}}
}%
{\put(6391,839){\line( 1, 0){450}}
\put(6841,839){\line( 0,-1){1305}}
\put(6841,-466){\line( 1, 0){450}}
\put(7291,-466){\line( 0, 1){855}}
\put(7291,389){\line( 1, 0){450}}
\put(7741,389){\line( 0, 1){450}}
\put(7741,839){\line( 1, 0){450}}
}%
{\put(6841,-511){\line( 0,-1){855}}
\put(6841,-1366){\line( 1, 0){450}}
\put(7291,-1366){\line( 0, 1){855}}
\put(7291,-511){\line(-1, 0){450}}
}%
{\put(4906,839){\line( 1, 0){450}}
\put(5356,839){\line( 0,-1){405}}
\put(5356,434){\line( 0,-1){ 45}}
\put(5356,389){\line( 1, 0){450}}
\put(5806,389){\line( 0, 1){450}}
\put(5806,839){\line( 1, 0){450}}
}%
{\put(5356,344){\line( 0,-1){855}}
\put(5356,-511){\line( 1, 0){450}}
\put(5806,-511){\line( 0, 1){855}}
\put(5806,344){\line(-1, 0){450}}
}%
{\put(5356,-556){\line( 0,-1){810}}
\put(5356,-1366){\line( 1, 0){450}}
\put(5806,-1366){\line( 0, 1){810}}
\put(5806,-556){\line(-1, 0){450}}
}%
{\put(3421,839){\line( 1, 0){450}}
\put(3871,839){\line( 0,-1){900}}
\put(3871,-61){\line( 1, 0){450}}
\put(4321,-61){\line( 0, 1){900}}
\put(4321,839){\line( 1, 0){450}}
}%
{\put(3871,-106){\line( 0,-1){1260}}
\put(3871,-1366){\line( 1, 0){450}}
\put(4321,-1366){\line( 0, 1){1215}}
\put(4321,-151){\line( 0, 1){ 45}}
\put(4321,-106){\line(-1, 0){450}}
}%
{\put(1486,839){\line( 1, 0){450}}
\put(1936,839){\line( 0,-1){2205}}
\put(1936,-1366){\line( 1, 0){450}}
\put(2386,-1366){\line( 0, 1){1755}}
\put(2386,389){\line( 1, 0){450}}
\put(2836,389){\line( 0, 1){450}}
\put(2836,839){\line( 1, 0){450}}
}%
{\put(10351,839){\line( 1, 0){450}}
\put(10801,839){\line( 0,-1){2205}}
\put(10801,-1366){\line( 1, 0){450}}
\put(11251,-1366){\line( 0, 1){855}}
\put(11251,-511){\line( 1, 0){405}}
\put(11656,-511){\line( 0, 1){450}}
\put(11656,-61){\line(-1, 0){405}}
\put(11251,-61){\line( 0, 1){900}}
\put(11251,839){\line( 1, 0){900}}
}%
{\multiput(271,-1411)(114.78261,0.00000){104}{\line( 1, 0){ 57.391}}
}%
\put(600,-1900){\footnotesize $1$}
\put(1956,-1900){\footnotesize $+4t$}
\put(3200,-1900){\footnotesize $-(n-1)t$}
\put(4800,-1900){\footnotesize $+\left(n-1\atop2\right)t^2$}
\put(6400,-1900){\footnotesize $-4(n-1)t^2$}
\put(8300,-1900){\footnotesize $+4(n-2)t^2$}
\put(10300,-1900){\footnotesize $+4(n-2)t^2$}
\end{picture}%

\vskip0.8cm
\begin{center}
\begin{minipage}{9cm}
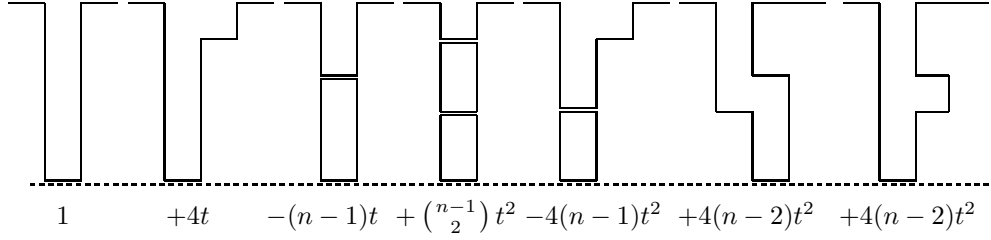
{\footnotesize{\centerline{Fig. 2. $P_n$: up to the $t^2$-terms
      dependent on $n$.}}} 
\end{minipage}
\end{center}
\vskip0.4cm
\begin{picture}(12174,2734)(-11,-1423)
\thinlines
{\put(6706,839){\line( 1, 0){450}}
\put(7156,839){\line( 0,-1){2205}}
\put(7156,-1366){\line( 1, 0){450}}
\put(7606,-1366){\line( 0, 1){1755}}
\put(7606,389){\line( 1, 0){450}}
\put(8056,389){\line( 0, 1){450}}
\put(8056,839){\line( 1, 0){450}}
}%
{\put(7156,839){\line( 2, 1){684}}
\put(7831,1199){\line( 1, 0){900}}
\put(8061,839){\line( 2, 1){684}}
\put(7156,839){\line( 1, 0){900}}
\put(7500,1020){\line( 1, 0){900}}
\put(7600,839){\line( 2, 1){684}}
}%
{\multiput(271,-1411)(114.78261,0.00000){104}{\line( 1, 0){ 57.391}}
}%
{\put(  1,839){\line( 1, 0){450}}
\put(451,839){\line( 0,-1){2205}}
\put(451,-1366){\line( 1, 0){450}}
\put(901,-1366){\line( 0, 1){1755}}
\put(901,389){\line( 1, 0){900}}
\put(1801,389){\line( 0, 1){450}}
\put(1801,839){\line( 1, 0){450}}
}%
{\put(2386,839){\line( 1, 0){450}}
\put(2836,839){\line( 0,-1){450}}
\put(2836,389){\line( 1, 0){450}}
\put(3286,389){\line( 0,-1){1755}}
\put(3286,-1366){\line( 1, 0){450}}
\put(3736,-1366){\line( 0, 1){1755}}
\put(3736,389){\line( 1, 0){450}}
\put(4186,389){\line( 0, 1){450}}
\put(4186,839){\line( 1, 0){450}}
}%
{\put(4771,839){\line( 1, 0){450}}
\put(5221,839){\line( 0,-1){2205}}
\put(5221,-1366){\line( 1, 0){450}}
\put(5671,-1366){\line( 0, 1){1305}}
\put(5671,-61){\line( 1, 0){450}}
\put(6121,-61){\line( 0, 1){900}}
\put(6121,839){\line( 1, 0){450}}
}%
{\put(8776,839){\line( 1, 0){450}}
\put(9226,839){\line( 0,-1){2205}}
\put(9226,-1366){\line( 1, 0){450}}
\put(9676,-1366){\line( 0, 1){2205}}
\put(9676,839){\line( 1, 0){450}}
}%
\thicklines
{\multiput(9226,839)(0.00000,128.57143){4}{\line( 0, 1){ 64.286}}
\multiput(9226,1289)(128.57143,0.00000){4}{\line( 1, 0){ 64.286}}
\multiput(9676,1289)(0.00000,-128.57143){4}{\line( 0,-1){ 64.286}}
}%
\put(400,-1900){\footnotesize $+12t^2$}
\put(3300,-1900){\footnotesize $+6t^2$}
\put(5300,-1900){\footnotesize $+4t^2$}
\put(7300,-1900){\footnotesize $+4t^2$}
\put(9300,-1900){\footnotesize $-6t^2$}
\end{picture}%

\vskip0.8cm
\begin{center}
\begin{minipage}{9cm}
{\footnotesize{\centerline{Fig. 3. $P_n$: $t^2$-terms independent of $n$.}}}
\end{minipage}
\end{center}

\vskip0.5cm
\begin{picture}(11904,2724)(-11,-1873)
\thinlines
{\multiput(  1,-1861)(114.78261,0.00000){104}{\line( 1, 0){ 57.391}}
}%
{\put(  1,839){\line( 1, 0){450}}
\put(451,839){\line( 0,-1){900}}
\put(451,-61){\line( 1, 0){450}}
\put(901,-61){\line( 0, 1){900}}
\put(901,839){\line( 1, 0){450}}
}%
{\put(451,-961){\line( 0,-1){855}}
\put(451,-1816){\line( 1, 0){450}}
\put(901,-1816){\line( 0, 1){855}}
\put(901,-961){\line(-1, 0){450}}
}%
{\put(10036,839){\line( 1, 0){450}}
\put(10486,839){\line( 0,-1){900}}
\put(10486,-61){\line( 1, 0){450}}
\put(10936,-61){\line( 0, 1){900}}
\put(10936,839){\line( 1, 0){900}}
}%
{\put(10981,389){\line( 0,-1){2205}}
\put(10981,-1816){\line( 1, 0){405}}
\put(11386,-1816){\line( 0, 1){2205}}
\put(11386,389){\line(-1, 0){405}}
}%
{\put(1846,839){\line( 1, 0){450}}
\put(2296,839){\line( 0,-1){1305}}
\put(2296,-466){\line( 1, 0){450}}
\put(2746,-466){\line( 0, 1){855}}
\put(2746,389){\line( 1, 0){450}}
\put(3196,389){\line( 0, 1){450}}
\put(3196,839){\line( 1, 0){450}}
}%
{\put(2296,-511){\line( 0,-1){855}}
\put(2296,-1366){\line( 1, 0){450}}
\put(2746,-1366){\line( 0, 1){855}}
\put(2746,-511){\line(-1, 0){450}}
}%
{\put(2296,-1411){\line( 0,-1){405}}
\put(2296,-1816){\line( 1, 0){450}}
\put(2746,-1816){\line( 0, 1){405}}
\put(2746,-1411){\line(-1, 0){450}}
}%
{\put(4006,839){\line( 1, 0){450}}
\put(4456,839){\line( 0,-1){1755}}
\put(4456,-916){\line( 1, 0){450}}
\put(4906,-916){\line( 0, 1){405}}
\put(4906,-511){\line( 1, 0){405}}
\put(5311,-511){\line( 0, 1){450}}
\put(5311,-61){\line(-1, 0){405}}
\put(4906,-61){\line( 0, 1){900}}
\put(4906,839){\line( 1, 0){900}}
}%
{\put(4456,-961){\line( 0,-1){855}}
\put(4456,-1816){\line( 1, 0){450}}
\put(4906,-1816){\line( 0, 1){855}}
\put(4906,-961){\line(-1, 0){450}}
}%
{\put(6166,839){\line( 1, 0){450}}
\put(6616,839){\line( 0,-1){1350}}
\put(6616,-511){\line( 1, 0){450}}
\put(7066,-511){\line( 0,-1){855}}
\put(7066,-1366){\line( 1, 0){450}}
\put(7516,-1366){\line( 0, 1){1305}}
\put(7516,-61){\line(-1, 0){450}}
\put(7066,-61){\line( 0, 1){900}}
\put(7066,839){\line( 1, 0){900}}
}%
{\put(7066,-1411){\line( 0,-1){405}}
\put(7066,-1816){\line( 1, 0){450}}
\put(7516,-1816){\line( 0, 1){405}}
\put(7516,-1411){\line(-1, 0){450}}
}%
{\put(8326,839){\line( 1, 0){450}}
\put(8776,839){\line( 0,-1){2655}}
\put(8776,-1816){\line( 1, 0){450}}
\put(9226,-1816){\line( 0, 1){2655}}
\put(9226,839){\line( 1, 0){450}}
}%
\thicklines
{\multiput(8731,-511)(0.00000,-180.00000){3}{\line( 0,-1){ 90.000}}
\multiput(8731,-961)(180.00000,0.00000){3}{\line( 1, 0){ 90.000}}
\multiput(9181,-961)(0.00000,180.00000){3}{\line( 0, 1){ 90.000}}
\multiput(9181,-511)(-180.00000,0.00000){3}{\line(-1, 0){ 90.000}}
}%
\thinlines
{\put(451,-106){\line( 0,-1){405}}
\put(451,-511){\line( 1, 0){450}}
\put(901,-511){\line( 0, 1){405}}
\put(901,-106){\line(-1, 0){450}}
}%
{\put(451,-556){\line( 0,-1){360}}
\put(451,-916){\line( 1, 0){450}}
\put(901,-916){\line( 0, 1){360}}
\put(901,-556){\line(-1, 0){450}}
}%
\put(0,-2400){\footnotesize $-\left(n-1\atop3\right)t^3$}
\put(1700,-2400){\footnotesize $+4\left(n-1\atop2\right)t^3$}
\put(3600,-2400){\footnotesize $-8\left(n-1\atop2\right)t^3$}
\put(6100,-2400){\footnotesize $-8\left(n-1\atop2\right)t^3$}
\put(8400,-2400){\footnotesize $-5(n-2)t^3$}
\put(10100,-2400){\footnotesize $-5(n-2)t^3$}
\end{picture}%
\vskip0.9cm
\begin{center}
\begin{minipage}{9cm}
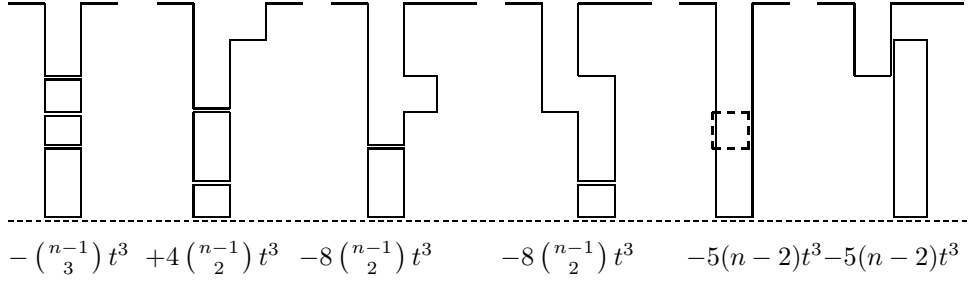
{\footnotesize{\centerline{Fig. 4. $P_n$: $t^3$-terms, 
cubic, quadratic (all) or linear (continued on next two Figs.) in $n$.}}} 
\end{minipage}
\end{center}
\vskip0.4cm
\begin{picture}(12174,2734)(-11,-1423)
\thinlines
{\multiput(271,-1411)(114.78261,0.00000){104}{\line( 1, 0){ 57.391}}
}%
\thicklines
{\multiput(9226,839)(0.00000,128.57143){4}{\line( 0, 1){ 64.286}}
\multiput(9226,1289)(128.57143,0.00000){4}{\line( 1, 0){ 64.286}}
\multiput(9676,1289)(0.00000,-128.57143){4}{\line( 0,-1){ 64.286}}
}%
\thinlines
{\put(  1,839){\line( 1, 0){450}}
\put(451,839){\line( 0,-1){1305}}
\put(451,-466){\line( 1, 0){450}}
\put(901,-466){\line( 0, 1){855}}
\put(901,389){\line( 1, 0){900}}
\put(1801,389){\line( 0, 1){450}}
\put(1801,839){\line( 1, 0){450}}
}%
{\put(451,-511){\line( 0,-1){855}}
\put(451,-1366){\line( 1, 0){450}}
\put(901,-1366){\line( 0, 1){855}}
\put(901,-511){\line(-1, 0){450}}
}%

{\put(7156,839){\line( 2, 1){684}}
\put(7831,1199){\line( 1, 0){900}}
\put(8061,839){\line( 2, 1){684}}
\put(7156,839){\line( 1, 0){900}}
\put(7500,1020){\line( 1, 0){900}}
\put(7600,839){\line( 2, 1){684}}
}%

{\put(2386,839){\line( 1, 0){450}}
\put(2836,839){\line( 0,-1){450}}
\put(2836,389){\line( 1, 0){450}}
\put(3286,389){\line( 0,-1){855}}
\put(3286,-466){\line( 1, 0){450}}
\put(3736,-466){\line( 0, 1){855}}
\put(3736,389){\line( 1, 0){450}}
\put(4186,389){\line( 0, 1){450}}
\put(4186,839){\line( 1, 0){450}}
}%
{\put(4771,839){\line( 1, 0){450}}
\put(5221,839){\line( 0,-1){1305}}
\put(5221,-466){\line( 1, 0){450}}
\put(5671,-466){\line( 0, 1){405}}
\put(5671,-61){\line( 1, 0){450}}
\put(6121,-61){\line( 0, 1){900}}
\put(6121,839){\line( 1, 0){450}}
}%
{\put(6706,839){\line( 1, 0){450}}
\put(7156,839){\line( 0,-1){1305}}
\put(7156,-466){\line( 1, 0){450}}
\put(7606,-466){\line( 0, 1){855}}
\put(7606,389){\line( 1, 0){450}}
\put(8056,389){\line( 0, 1){450}}
\put(8056,839){\line( 1, 0){450}}
}%
{\put(8776,839){\line( 1, 0){450}}
\put(9226,839){\line( 0,-1){1305}}
\put(9226,-466){\line( 1, 0){450}}
\put(9676,-466){\line( 0, 1){1305}}
\put(9676,839){\line( 1, 0){450}}
}%
{\put(3286,-511){\line( 0,-1){855}}
\put(3286,-1366){\line( 1, 0){450}}
\put(3736,-1366){\line( 0, 1){855}}
\put(3736,-511){\line(-1, 0){450}}
}%
{\put(5221,-511){\line( 0,-1){855}}
\put(5221,-1366){\line( 1, 0){450}}
\put(5671,-1366){\line( 0, 1){855}}
\put(5671,-511){\line(-1, 0){450}}
}%
{\put(7156,-511){\line( 0,-1){855}}
\put(7156,-1366){\line( 1, 0){450}}
\put(7606,-1366){\line( 0, 1){855}}
\put(7606,-511){\line(-1, 0){450}}
}%
{\put(9226,-511){\line( 0,-1){855}}
\put(9226,-1366){\line( 1, 0){450}}
\put(9676,-1366){\line( 0, 1){855}}
\put(9676,-511){\line(-1, 0){450}}
}%
\put(100,-1900){\footnotesize $-12(n-1)t^3$}
\put(2700,-1900){\footnotesize $-6(n-1)t^3$}
\put(4800,-1900){\footnotesize $-4(n-2)t^3$}
\put(6800,-1900){\footnotesize $-4(n-1)t^3$}
\put(8800,-1900){\footnotesize $+6(n-1)t^3$}
\end{picture}%

\vskip0.9cm
\begin{center}
\begin{minipage}{9cm}
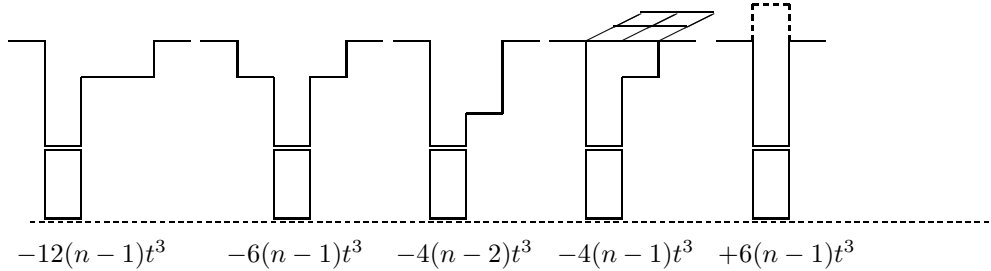
{\footnotesize{\centerline{Fig. 5. $P_n$: $t^3$-terms, analog of $t^2$
    terms on Fig. 3.}}} 
\end{minipage}
\end{center}
\vskip1cm
\begin{picture}(11364,2274)(-11,-1423)
\thinlines
{\multiput(  1,-1411)(113.96985,0.00000){100}{\line( 1, 0){ 56.985}}
}%
{\put(  1,839){\line( 1, 0){450}}
\put(451,839){\line( 0,-1){1800}}
\put(451,-961){\line( 1, 0){450}}
\put(901,-961){\line( 0,-1){405}}
\put(901,-1366){\line( 1, 0){450}}
\put(1351,-1366){\line( 0, 1){855}}
\put(1351,-511){\line(-1, 0){450}}
\put(901,-511){\line( 0, 1){900}}
\put(901,389){\line( 1, 0){450}}
\put(1351,389){\line( 0, 1){450}}
\put(1351,839){\line( 1, 0){450}}
}%
{\put(4951,839){\line( 1, 0){450}}
\put(5401,839){\line( 0,-1){2205}}
\put(5401,-1366){\line( 1, 0){450}}
\put(5851,-1366){\line( 0, 1){855}}
\put(5851,-511){\line( 1, 0){450}}
\put(6301,-511){\line( 0, 1){450}}
\put(6301,-61){\line(-1, 0){450}}
\put(5851,-61){\line( 0, 1){450}}
\put(5851,389){\line( 1, 0){450}}
\put(6301,389){\line( 0, 1){450}}
\put(6301,839){\line( 1, 0){450}}
}%
{\put(2701,839){\line( 1, 0){450}}
\put(3151,839){\line( 0,-1){1800}}
\put(3151,-961){\line( 1, 0){450}}
\put(3601,-961){\line( 0,-1){405}}
\put(3601,-1366){\line( 1, 0){450}}
\put(4051,-1366){\line( 0, 1){1305}}
\put(4051,-61){\line(-1, 0){450}}
\put(3601,-61){\line( 0, 1){900}}
\put(3601,839){\line( 1, 0){450}}
}%
{\put(7651,839){\line( 1, 0){450}}
\put(8101,839){\line( 0,-1){2205}}
\put(8101,-1366){\line( 1, 0){450}}
\put(8551,-1366){\line( 0, 1){405}}
\put(8551,-961){\line( 1, 0){450}}
\put(9001,-961){\line( 0, 1){855}}
\put(9001,-106){\line(-1, 0){450}}
\put(8551,-106){\line( 0, 1){945}}
\put(8551,839){\line( 1, 0){450}}
\put(9001,839){\line( 0, 1){  0}}
}%
\put(100,-1900){\footnotesize $+(16(n-2)-4)t^3$}
\put(2700,-1900){\footnotesize $+4(n-3)t^3$}
\put(4800,-1900){\footnotesize $+(16(n-2)-4)t^3$}
\put(7400,-1900){\footnotesize $+4(n-3)t^3$}
\end{picture}%

\vskip0.9cm
\begin{center}
\begin{minipage}{9cm}
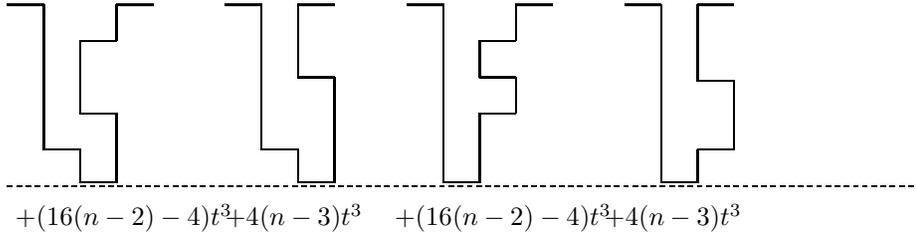
{\footnotesize{\centerline{Fig. 6. $P_n$: $t^3$-terms, 
linear in $n$ (continued from previous Figs.).}}} 
\end{minipage}
\end{center}
\vskip1cm
The leading terms up to $t^3$ and the double leg in $Q_n=Q_n^1+Q_n^2$
are shown on Fig. 7. 

\bigskip
\begin{picture}(11004,2274)(259,-1423)
\thinlines
{\put(451,839){\line( 1, 0){450}}
\put(901,839){\line( 0,-1){2205}}
\put(901,-1366){\line( 1, 0){900}}
\put(1801,-1366){\line( 0, 1){405}}
\put(1801,-961){\line(-1, 0){450}}
\put(1351,-961){\line( 0, 1){1800}}
\put(1351,839){\line( 1, 0){450}}
}%
{\put(2251,839){\line( 1, 0){450}}
\put(2701,839){\line( 0,-1){900}}
\put(2701,-61){\line( 1, 0){450}}
\put(3151,-61){\line( 0, 1){900}}
\put(3151,839){\line( 1, 0){450}}
}%
{\put(2701,-106){\line( 0,-1){1260}}
\put(2701,-1366){\line( 1, 0){900}}
\put(3601,-1366){\line( 0, 1){405}}
\put(3601,-961){\line(-1, 0){450}}
\put(3151,-961){\line( 0, 1){855}}
\put(3151,-106){\line(-1, 0){450}}
}%
{\put(4051,839){\line( 1, 0){450}}
\put(4501,839){\line( 0,-1){2205}}
\put(4501,-1366){\line( 1, 0){900}}
\put(5401,-1366){\line( 0, 1){405}}
\put(5401,-961){\line(-1, 0){450}}
\put(4951,-961){\line( 0, 1){1350}}
\put(4951,389){\line( 1, 0){450}}
\put(5401,389){\line( 0, 1){450}}
\put(5401,839){\line( 1, 0){450}}
}%
{\put(6301,839){\line( 1, 0){450}}
\put(6751,839){\line( 0,-1){2205}}
\put(6751,-1366){\line( 1, 0){900}}
\put(7651,-1366){\line( 0, 1){855}}
\put(7651,-511){\line(-1, 0){450}}
\put(7201,-511){\line( 0, 1){1350}}
\put(7201,839){\line( 1, 0){450}}
}%
{\put(9001,839){\line( 1, 0){450}}
\put(9451,839){\line( 0,-1){2205}}
\put(9451,-1366){\line( 1, 0){450}}
\put(9901,-1366){\line( 0, 1){2205}}
\put(9901,839){\line( 1, 0){450}}
}%
{\put(9946,-1366){\framebox(405,405){}}
}%
{\multiput(271,-1411)(130,0.00000){97}{\line( 1, 0){56.891}}
}%
{\put(11001,839){\line( 1, 0){450}}
\put(11451,839){\line( 0,-1){2205}}
\put(11451,-1366){\line( 1, 0){900}}
\put(12301,-1366){\line( 0, 1){2205}}
\put(12301,839){\line( 1, 0){450}}
}%
\put(1000,-1900){\footnotesize $4t^2$}
\put(2200,-1900){\footnotesize $-4(n-1)t^3$}
\put(4400,-1900){\footnotesize $+16t^3$}
\put(6800,-1900){\footnotesize $+4t^3$}
\put(9600,-1900){\footnotesize $-5t^3$}
\put(11600,-1900){\footnotesize $+2t^n$}
\end{picture}%
\vskip0.9cm
\begin{center}
\begin{minipage}{9cm}
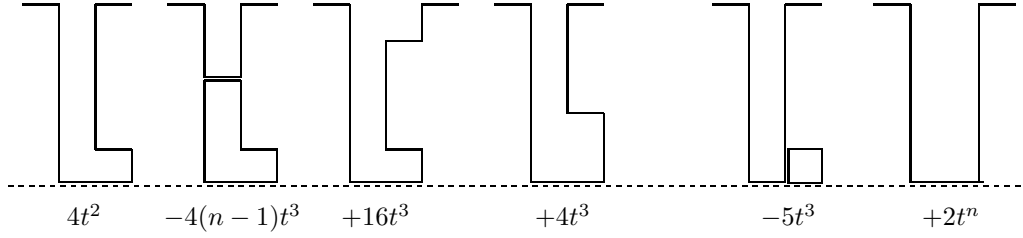
{\footnotesize{\centerline{Fig. 7. $Q_n=Q_n^1+Q_n^2$: up to order
      $t^3$, and double leg.}}} 
\end{minipage}
\end{center}

\bigskip

Formulas (\ref{cn}) and (\ref{an}) for $c_n$ and $a_n$ are consistent
with the recursion relations, notably the equation giving $P_{n+1}$
not used so far, implying 
\beqa
c_{n+1}&=&c_n+n+3\cr
a_{n+1}&=&a_n+c_n+8n-30\,.
\eeqa
For later purposes we note that
\beqa
a_n-a_{n-1}&=&\12(n-1)(n-2)+11n-32\cr
a_n-a_{n-1}+2c_n-4c_{n-1}+c_{n-2}&=&9n-35
\eeqa
Putting together (\ref{APQRS}), (\ref{induc}) and assuming $u=\O(t^2)$ gives 
for $n\ge3$
\begin{align}\label{AAA}
A_n&(u)-A_n(0)-t^2A_{n+1}(u)\\
&= \bigl(e^u-1-e^ut^2+e^ut^3-4e^{2u}t^4+9e^{2u}t^5\notag \\
&\qquad \qquad-5e^{2u}t^6-18e^{3u}t^6+66e^{3u}t^7-4e^{4u}t^7-139t^8+Ct^9\bigr)P_n\notag \\
&\ +\bigl(e^{2u}-1-2e^ut^2+2e^ut^3\bigr)Q_n
  +\bigl(-e^{2u}(t^3+6t^4)+11t^5-28t^6+Ct^7\bigr)Q_n^1\notag \\
&\quad+(-9t^4+20t^5+Ct^6)Q_n^2+\bigl(e^{3u}-1-3e^ut^2+3e^ut^3\bigr)R_n\notag \\
&\quad-\bigl(2t^3+10t^4+Ct^5\bigr)R_n^1-(t^3+Ct^4)R_n^2\notag \\
&\quad+(e^{4u}-1-4e^ut^2+4e^ut^3+Ct^4)S_n-(4t^3+Ct^4)S_n\notag \\
&\quad+(e^{5u}-1-5e^ut^2)T_n+\O(t^{10}).\notag
\end{align}
Contributions from $U_n$ have been absorbed into $\O(t^{10})$,
thanks to $u=\O(t^2)$.
For $n=2$, (\ref{AAA}) is valid up to order $t^6$, with an error $\O(t^7)$.
By \eqref{PQRS}, in each of the expansions $P_n,Q_n^1,Q_n^2$, etc., $n$-dependent terms only appear at 
one or more orders less in $t$ than the largest-order term, 
and when \eqref{AAA} is multiplied out, 
the unspecified constants $C$ only appear at order 
$t^9$ or less.  Therefore 
the constants $C$ appear in $n$-dependent terms
only at order $t^{10}$ or less, so that, while the constants $C$ are relevant to the value of $B_9$ in 
\eqref{123456}, they are not relevant to establishing that $B_9$ is $n$-independent.
The constants $C$ thus play the role of placeholders, permitting the $\O(t^{10})$ error 
term which allows analysis of the dependence of $B_9$ on $n$.

\bigskip\noindent{\bf First excitations in $B_n(u)$:}
\begin{align}\label{tiPQRS}
\ti P_n&=t-(n-1)t^2+\ti c_nt^3-\ti a_nt^4+\ti d_nt^5+\O(t^6)\,,\qquad n\ge4\notag\\
\ti Q_n^1&=4\bigl[t^3-(n-2)t^4+(\ti c_{n-1}+7)t^5 + (-\ti a_{n-1} - \ti c_{n-1} + \ti c_{n-2} -6n+17)t^6\bigr]\notag \\
&\qquad+2t^{n+2}+\O(t^7)\,,\qquad n\ge4\notag\\
\ti Q_n^2&=-5t^4+5(n-1)t^5+\O(t^6)\,,\qquad n\ge2\notag\\
\ti R_n^1&=18t^5-(18n-42)t^6+\O(t^7)\,,\qquad n\ge2\notag\\
\ti R_n^2 &= -48t^6 + \O(t^7)\,,\qquad n \ge 2,\cr
\end{align}
with
\beqa\label{ticn}
\ti c_n&=&{{n-1} \choose 2}+8n-13\\\label{tian}
\ti a_n&=&{{n-1} \choose 3}+16{{n-1} \choose 2}+5n+2\\
\ti d_n&=&{{n-1}\choose 4} + 24{{n-1} \choose 3} + 79{{n-1} \choose 2} - 31n + 75.\label{tidn}
\eeqa
The expansion for $\ti P_n$ is valid for $n=3$ with one order less, and
for $n=2$ with two orders less, and for $n=1$ with three orders less.
The expansion for $\ti Q_n$ is valid for $n=3$ with one order less, and
for $n=2$ with two orders less.
Formula (\ref{ticn}) was obtained using Fig. 8. The last two diagrams in Fig. 8
belong to the second term inside the parentheses, 4th term in (\ref{5term}),
defining $\ti P_n$. Formula (\ref{tian}) was obtained using the last
diagram in Fig. 8 and all diagrams in Fig.~9. 
Formula (\ref{tidn}) was obtained using the diagrams in Fig.~15.
\vskip0.6cm
\begingroup\makeatletter\ifx\SetFigFont\undefined%
\gdef\SetFigFont#1#2#3#4#5{%
  \reset@font\fontsize{#1}{#2pt}%
  \fontfamily{#3}\fontseries{#4}\fontshape{#5}%
  \selectfont}%
\fi\endgroup%
\begin{picture}(11724,2724)(-11,-1423)
\thinlines
{\put(451,794){\line( 0,-1){2160}}
\put(451,-1366){\line( 1, 0){450}}
\put(901,-1366){\line( 0, 1){2160}}
\put(901,794){\line(-1, 0){450}}
}%
{\put(1801,-106){\line( 0,-1){1260}}
\put(1801,-1366){\line( 1, 0){450}}
\put(2251,-1366){\line( 0, 1){1215}}
\put(2251,-151){\line( 0, 1){ 45}}
\put(2251,-106){\line(-1, 0){450}}
}%
{\put(1801,794){\line( 0,-1){855}}
\put(1801,-61){\line( 1, 0){450}}
\put(2251,-61){\line( 0, 1){855}}
\put(2251,794){\line(-1, 0){450}}
}%
{\multiput(  1,839)(114.14634,0.00000){103}{\line( 1, 0){ 57.073}}}
{\multiput(  1,-1411)(114.14634,0.00000){103}{\line( 1, 0){ 57.073}}}
{\put(10351,-106){\line( 0,-1){1260}}
\put(10351,-1366){\line( 1, 0){450}}
\put(10801,-1366){\line( 0, 1){1215}}
\put(10801,-151){\line( 0, 1){ 45}}
\put(10801,-106){\line(-1, 0){450}}
}%
{\put(10351,1289){\line( 0,-1){1350}}
\put(10351,-61){\line( 1, 0){450}}
\put(10801,-61){\line( 0, 1){1350}}
\put(10801,1289){\line(-1, 0){450}}
}%
{\put(9001,1289){\line( 0,-1){2655}}
\put(9001,-1366){\line( 1, 0){450}}
\put(9451,-1366){\line( 0, 1){2160}}
\put(9451,794){\line( 0, 1){495}}
\put(9451,1289){\line(-1, 0){450}}
}%
{\put(6976,794){\line( 0,-1){2160}}
\put(6976,-1366){\line( 1, 0){450}}
\put(7426,-1366){\line( 0, 1){855}}
\put(7426,-511){\line( 1, 0){405}}
\put(7831,-511){\line( 0, 1){450}}
\put(7831,-61){\line(-1, 0){405}}
\put(7426,-61){\line( 0, 1){855}}
\put(7426,794){\line(-1, 0){450}}
}%
{\put(4951,794){\line( 0,-1){1305}}
\put(4951,-511){\line( 1, 0){450}}
\put(5401,-511){\line( 0,-1){855}}
\put(5401,-1366){\line( 1, 0){450}}
\put(5851,-1366){\line( 0, 1){1305}}
\put(5851,-61){\line(-1, 0){450}}
\put(5401,-61){\line( 0, 1){855}}
\put(5401,794){\line(-1, 0){450}}
}%
{\put(3376,434){\framebox(450,360){}}
}%
{\put(3376,389){\line( 0,-1){855}}
\put(3376,-466){\line( 1, 0){450}}
\put(3826,-466){\line( 0, 1){855}}
\put(3826,389){\line(-1, 0){450}}
}%
{\put(3376,-511){\line( 0,-1){855}}
\put(3376,-1366){\line( 1, 0){450}}
\put(3826,-1366){\line( 0, 1){855}}
\put(3826,-511){\line(-1, 0){450}}
}%
\put(600,-1900){\footnotesize $t$}
\put(1200,-1900){\footnotesize $-(n-1)t^2$}
\put(2800,-1900){\footnotesize $+\left(n-1\atop2\right)t^3$}
\put(4550,-1900){\footnotesize $+4(n-2)t^3$}
\put(6550,-1900){\footnotesize $+4(n-1)t^3$}
\put(9000,-1900){\footnotesize $-t^3$}
\put(10000,-1900){\footnotesize $+(n-1)t^4$}
\end{picture}%
\vskip0.8cm
\begin{center}
\begin{minipage}{9cm}
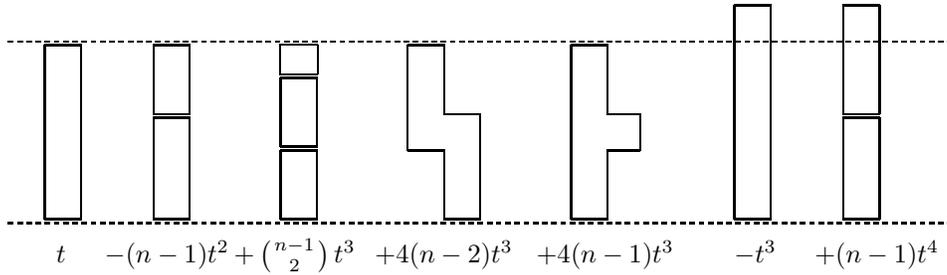
{\footnotesize{\centerline{Fig. 8. $\ti P_n$: $t^3$-terms, 
and $t^4$-term incompatible with $I_{n+1}$.}}} 
\end{minipage}
\end{center}

\vskip1cm
\begingroup\makeatletter\ifx\SetFigFont\undefined%
\gdef\SetFigFont#1#2#3#4#5{%
  \reset@font\fontsize{#1}{#2pt}%
  \fontfamily{#3}\fontseries{#4}\fontshape{#5}%
  \selectfont}%
\fi\endgroup%
\begin{picture}(11904,2679)(-11,-1873)
\thinlines
{\multiput(  1,839)(114.14634,0.00000){103}{\line( 1, 0){ 57.073}}}
{\multiput(  1,-1861)(114.78261,0.00000){104}{\line( 1, 0){ 57.391}}
}%
{\put(451,-961){\line( 0,-1){855}}
\put(451,-1816){\line( 1, 0){450}}
\put(901,-1816){\line( 0, 1){855}}
\put(901,-961){\line(-1, 0){450}}
}%
{\put(451,-106){\line( 0,-1){405}}
\put(451,-511){\line( 1, 0){450}}
\put(901,-511){\line( 0, 1){405}}
\put(901,-106){\line(-1, 0){450}}
}%
{\put(451,-556){\line( 0,-1){360}}
\put(451,-916){\line( 1, 0){450}}
\put(901,-916){\line( 0, 1){360}}
\put(901,-556){\line(-1, 0){450}}
}%
{\put(451,794){\line( 0,-1){855}}
\put(451,-61){\line( 1, 0){450}}
\put(901,-61){\line( 0, 1){855}}
\put(901,794){\line(-1, 0){450}}
}%
{\put(10801,794){\line( 0,-1){2610}}
\put(10801,-1816){\line( 1, 0){450}}
\put(11251,-1816){\line( 0, 1){855}}
\put(11251,-961){\line( 1, 0){450}}
\put(11701,-961){\line( 0, 1){900}}
\put(11701,-61){\line(-1, 0){450}}
\put(11251,-61){\line( 0, 1){855}}
\put(11251,794){\line(-1, 0){450}}
}%
{\put(8776,794){\line( 0, 1){  0}}
\put(8776,794){\line( 0,-1){1755}}
\put(8776,-961){\line( 1, 0){450}}
\put(9226,-961){\line( 0,-1){855}}
\put(9226,-1816){\line( 1, 0){450}}
\put(9676,-1816){\line( 0, 1){1755}}
\put(9676,-61){\line(-1, 0){450}}
\put(9226,-61){\line( 0, 1){855}}
\put(9226,794){\line(-1, 0){450}}
}%
{\put(7201,794){\line( 0,-1){855}}
\put(7201,-61){\line( 1, 0){450}}
\put(7651,-61){\line( 0, 1){855}}
\put(7651,794){\line(-1, 0){450}}
}%
{\put(7696,389){\line( 0,-1){2205}}
\put(7696,-1816){\line( 1, 0){405}}
\put(8101,-1816){\line( 0, 1){2205}}
\put(8101,389){\line(-1, 0){405}}
}%
\thicklines
{\multiput(5671,-511)(0.00000,-180.00000){3}{\line( 0,-1){ 90.000}}
\multiput(5671,-961)(180.00000,0.00000){3}{\line( 1, 0){ 90.000}}
\multiput(6121,-961)(0.00000,180.00000){3}{\line( 0, 1){ 90.000}}
\multiput(6121,-511)(-180.00000,0.00000){3}{\line(-1, 0){ 90.000}}
}%
\thinlines
{\put(5626,794){\line( 0, 1){  0}}
\put(5626,794){\line( 0,-1){2610}}
\put(5626,-1816){\line( 1, 0){450}}
\put(6076,-1816){\line( 0, 1){2610}}
\put(6076,794){\line(-1, 0){450}}
}%
{\put(2026,794){\line( 0,-1){1710}}
\put(2026,-916){\line( 1, 0){450}}
\put(2476,-916){\line( 0, 1){405}}
\put(2476,-511){\line( 1, 0){405}}
\put(2881,-511){\line( 0, 1){450}}
\put(2881,-61){\line(-1, 0){405}}
\put(2476,-61){\line( 0, 1){855}}
\put(2476,794){\line(-1, 0){450}}
}%
{\put(2026,-961){\line( 0,-1){855}}
\put(2026,-1816){\line( 1, 0){450}}
\put(2476,-1816){\line( 0, 1){855}}
\put(2476,-961){\line(-1, 0){450}}
}%
{\put(3826,794){\line( 0, 1){  0}}
\put(3826,794){\line( 0,-1){1305}}
\put(3826,-511){\line( 1, 0){450}}
\put(4276,-511){\line( 0,-1){855}}
\put(4276,-1366){\line( 1, 0){450}}
\put(4726,-1366){\line( 0, 1){1305}}
\put(4726,-61){\line(-1, 0){450}}
\put(4276,-61){\line( 0, 1){855}}
\put(4276,794){\line(-1, 0){450}}
}%
{\put(4276,-1411){\line( 0,-1){405}}
\put(4276,-1816){\line( 1, 0){450}}
\put(4726,-1816){\line( 0, 1){405}}
\put(4726,-1411){\line(-1, 0){450}}
}%
\put(-50,-2400){\footnotesize $-\left(n-1\atop3\right)t^4$}
\put(1500,-2400){\footnotesize $-4(n-1)^2t^4$}
\put(3400,-2400){\footnotesize $-8\left(n-1\atop2\right)t^4$}
\put(5100,-2400){\footnotesize $-5(n-1)t^4$}
\put(6800,-2400){\footnotesize $-5(n-2)t^4$}
\put(8500,-2400){\footnotesize $+4(n-3)t^4$}
\put(10200,-2400){\footnotesize $+4(n-2)t^4$}
\end{picture}%
\vskip0.9cm
\begin{center}
\begin{minipage}{9cm}
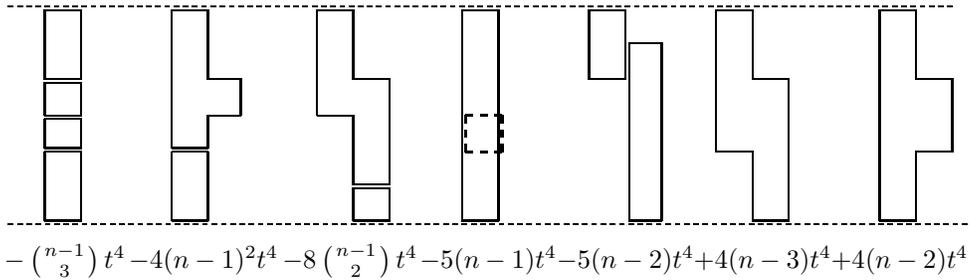
{\footnotesize{\centerline{Fig. 9. $\ti P_n$: $t^4$-terms, other than
Fig. 8.}}} 
\end{minipage}
\end{center}
\vskip1cm

These $\ti c_n$ and $\ti a_n$ are consistent with the recursion
relations, which imply 
\begin{align}
\ti c_{n+1}&=\ti c_n+n+7,\notag\\
\ti a_{n+1}&=\ti a_n+\ti c_n+8n+2.
\end{align}
Then
\begin{align}
\ti c_n-\ti c_{n-1}&=n+6,\notag\\
\ti a_{n+2}-\ti a_n&=(n-1)(n-2)+33n-7,
\end{align}
and from this we obtain
\beq\label{tianrec}
\ti a_{n+2}-\ti a_n+2\ti c_n-4\ti c_{n-1}=21n+43.
\eeq
Putting together (\ref{BPQRS}) and the analog of (\ref{induc}) for $\ti P_n, \ti Q_n$, etc., and
assuming $u=\O(t^2)$, gives for $n\ge2$
\begin{align}\label{BBB}
B_n&(u)-B_n(0)-t^2B_{n+1}(0)\notag \\
&=\bigl(e^u-1-t^2+t^3-4t^4+9t^5-23t^6+62t^7-139t^8\bigr)\ti P_n\notag \\
&\quad+\bigl(e^{2u}-1-2t^2+2t^3\bigr)\ti Q_n+\bigl(-t^3-6t^4+11t^5-28t^6\bigr)\ti Q_n^1\notag \\
&\quad-(9t^4-20t^5)\ti Q_n^2+\bigl(e^{3u}-1-3t^2+3t^3\bigr)\ti R_n-(2t^3+10t^4)\ti R_n^1 \notag \\
&\quad-t^3\ti R_n^2+\O(t^{10}),
\end{align}
while from \eqref{BPQRS} and \eqref{tiPQRS},
\begin{align}\label{B2}
t^4B_{n+2}(0)&=t^5-(n+1)t^6+(\ti c_{n+2}+4)t^7
-\bigl(\ti a_{n+2}+4n+5\bigr)t^8\notag \\
&\quad+\big( \ti d_{n+2} + 4\ti c_{n+1} + 5n + 41 \big)t^9+\O(t^{10}),
\end{align}
\beq\label{B3}
t^6B_{n+3}(0)=t^7-(n+2)t^8+\big( \ti c_{n+3} + 4\big)t^9 + \O(t^{10}),
\eeq
\beq\label{B4}
t^8B_{n+4}(0)=t^9 + \O(t^{10}).
\eeq


\bigskip\noindent{\bf Assumption:}
\begin{align}
u=&-\ln(1-t^2)+b_3t^3+\dots+b_8t^8+b_9t^9+\O(t^{10})\notag\\
=&t^2+b_3t^3+(b_4+\12)t^4+b_5t^5+(b_6+{\text\frac{1}{3}})t^6+b_7t^7
+(b_8+{\text\frac{1}{4}})t^8+b_9t^9+\O(t^{10})\cr
\end{align}
\smallskip
Then, with $b_3=-1$ and $b_4=4$ where $b_3$ and $b_4$ don't appear explicitly,
\begin{align}\label{Aeu}
e^u(1-t^2)-1+e^ut^3&=(b_3+1)t^3+b_4t^4+(b_5+1)t^5+\O(t^6)\notag\\
&=4t^4+(b_5+1)t^5+(b_6-{\text\frac{1}{2}})t^6+(b_7+1)t^7\notag\\
&\qquad+(b_8+7)t^8+(b_9 + C)t^9+\O(t^{10})\notag\\
e^{2u}-1-2e^u(t^2-t^3)&=2(b_3+1)t^3+(2b_4+1)t^4+\O(t^5)\notag\\
&=9t^4+2b_5t^5+(2b_6+10)t^6+\O(t^7)\notag\\
e^{3u}-1-3e^u(t^2-t^3)&=3(b_3+1)t^3+3(b_4+1)t^4+\O(t^5)\notag\\
e^{4u}-1-4e^u(t^2-t^3)&=4(b_3+1)t^3+(4b_4+18)t^4+\O(t^5)
\end{align}
and
\begin{align}\label{Beu}
e^{u}-1-t^2+t^3&=(b_3+1)t^3+(b_4+1)t^4+\O(t^5)\notag\\
&=5t^4+(b_5-1)t^5+(b_6+{\text\frac{11}{2}})t^6
+(b_7+b_5-5)t^7\notag\\
&\qquad +(-b_5+b_6+b_8+\text\frac{27}{2})t^8+\O(t^9)\notag\\
e^{2u}-1-2t^2+2t^3&=2(b_3+1)t^3+(2b_4+3)t^4+\O(t^5)\notag\\
&=11t^4+(2b_5-4)t^5+(2b_6+22)t^6+\O(t^7)\notag\\
\quad e^{3u}-1-3t^2+3t^3&=18t^4+\O(t^5).
\end{align}
Then for $n\ge2$, from (\ref{AAA}), (\ref{Aeu}), (\ref{PQRS}),
\begin{align}\label{AAA3}
A_n(u)-A_n(0)-t^2A_{n+1}(u)&=\bigl[(b_3+1)t^3+(b_4-4)t^4\bigr]P_n+\O(t^5)\notag \\
=&(b_3+1)t^3+\bigl[b_4-4-(b_3+1)(n-5)\bigr]t^4+\O(t^5),
\end{align}
while from (\ref{BBB}), \eqref{B2}, (\ref{Beu}), (\ref{tiPQRS}),
\begin{align}
B_n(u)-\sum_{m=0}^\infty
t^{2m}B_{n+m}(0)&=\bigl[(b_3+1)t^3+(b_4-3)t^4\bigr]\ti P_n-t^5
+\O(t^6)\notag \\
&=(b_3+1)t^4+\bigl[b_4-4-(b_3+1)(n-1)\bigr]t^5+\O(t^6),
\end{align}
giving
\beq\label{fn3}
t^{-2n}(f_{n+1}-f_n)=(b_3+1)t^3+\bigl[b_4-4-(b_3+1)(n-4)\bigr]t^4+\O(t^5).
\eeq        
If $b_3=-1$, then from (\ref{AAA}), (\ref{Aeu}), (\ref{PQRS}), still for $n\ge2$,
\begin{align}\label{AAA4}
A_n(u)&-A_n(0)-t^2A_{n+1}(u)\notag \\
&=\bigl[(b_4-4)t^4+(b_5+10)t^5\bigr]P_n
-t^3Q_n+\O(t^6)\notag\\ 
&=(b_4-4)t^4+\bigl[b_5+6-(b_4-4)(n-5)\bigr]t^5-2t^{n+3}+\O(t^6),
\end{align}
giving
\beq\label{fn4}
t^{-2n}(f_{n+1}-f_n)=(b_4-4)t^4+\bigl[b_5+6-(b_4-4)(n-4)\bigr]t^5
-2t^{n+3}+\O(t^6).
\eeq        
If $b_4=4$, then from (\ref{AAA}), (\ref{Aeu}), (\ref{PQRS}), now for $n\ge3$,
\begin{align}\label{AAA5}
A_n&(u)-A_n(0)-t^2A_{n+1}(u)\notag\\
&=
  \bigl[(b_5+10)t^5+(b_6-\12-31)t^6\bigr]P_n+9t^4Q_n-(t^3+6t^4)Q_n^1+\O(t^7)\notag\\
&=(b_5+6)t^5+\bigl[b_6-\472-(b_5+6)(n-5)\bigr]t^6-2t^{n+3}+\O(t^7)
\end{align}
while from (\ref{BBB}), (\ref{Beu}), (\ref{tiPQRS}),
\beqa
B_n(u)-\sum_{m=0}^\infty t^{2m}B_{n+m}(0)
&=&[t^4+(b_5+8)t^5]\ti P_n-t^3\ti Q_n-t^5+(n+1)t^6+\O(t^7)\cr 
&=&(b_5+6)t^6+\O(t^7)
\eeqa
giving
\beq\label{fn5}
t^{-2n}(f_{n+1}-f_n)=(b_5+6)t^5+\bigl[b_6-\472-(b_5+6)(n-4)\bigr]t^6
-2t^{n+3}+\O(t^7)
\eeq        
If $b_5=-6$, then from (\ref{AAA}), (\ref{Aeu}), (\ref{PQRS}), now for $n\ge4$,
\begin{align}\label{AAA6}
A_n&(u)-A_n(0)-t^2A_{n+1}(u)\notag \\
&=
  \bigl[4t^5+(b_6-\12-31)t^6+(b_7+89)t^7\bigr]P_n+\bigl[9t^4-12t^5\bigr]Q_n\notag\\
&\qquad-(t^3+6t^4-9t^5)Q_n^1-9t^4Q_n^2-2t^3R_n^1+\O(t^8)\notag\\
&=(b_6-\472)t^6+\bigl[b_7+4(c_n-c_{n-1}-3n)-(b_6 -\652)(n-5)+85\bigr]t^7\cr
&\qquad-2t^{n+3}+\O(t^8)\notag\\
&=(b_6-\472)t^6+\bigl[b_7+53-(b_6 - \472)(n-5)\bigr]t^7-2t^{n+3}+\O(t^8),\cr
\end{align}
while from (\ref{BBB}), (\ref{Beu}), (\ref{tiPQRS}),
\begin{align}
B_n&(u)-B_n(0)-t^2B_{n+1}(0)\notag\\
&=\bigl[t^4+2t^5+(b_6-\352)t^6+(b_7+51)t^7\bigr]\ti P_n+(11t^4-16t^5)\ti Q_n\notag\\
&\qquad+(-t^3-6t^4+11t^5)\ti Q_n^1-9t^4\ti Q_n^2-2t^3\ti R_n+\O(t^9)\notag\\
&=t^5-(n+1)t^6+\bigl[\ti c_n+2n+b_6-\72\bigr]t^7\notag\\
&\qquad+\bigl[b_7-\ti a_n+2\ti c_n-4\ti c_{n-1}-(b_6+\52)n
+b_6-{\frac{41}{2}}\bigr]t^8+\O(t^9)
\end{align}
so that, with (\ref{B2}), (\ref{B3}), (\ref{tianrec}),
\begin{align}\label{BBBB}
B_n&(u)-\sum_{m=0}^\infty t^{2m}B_{n+m}(0)\notag\\
&=(b_6-\472)t^7
+\bigl[b_7+b_6+\ti a_{n+2}-\ti a_n+2\ti c_n-4\ti c_{n-1}-(b_6-\52)n
-{\frac{27}{2}}\bigr]t^8\notag\\
&\qquad+\O(t^9)\notag\\
&=(b_6-\472)t^7+\bigl[b_7+53-(b_6-\472)(n-1)\bigr]t^8+\O(t^9),\cr
\end{align}
giving
\beq\label{fn6}
t^{-2n}(f_{n+1}-f_n)=(b_6-\472)t^6+\bigl(b_7+53-(b_6-\472)(n-4)\bigr)t^7-2t^{n+3}
+\O(t^8).
\eeq        
If $b_6=\472$, then from (\ref{AAA}), (\ref{Aeu}), (\ref{PQRS}),
now for $n\ge5$, 
\begin{align}
A_n&(u)-A_n(0)-t^2A_{n+1}(u)\notag\\
&=
  \bigl[4t^5-8t^6+(b_7+89)t^7+(b_8-258)t^8\bigr]P_n 
  +\bigl(-t^3+3t^4-3t^5+19t^6\bigr)Q_n^1\notag\\
&\qquad+8t^5Q_n^2+\bigl(-2t^3+5t^4\bigr)R_n^1-t^3R_n^2-4t^3S_n+\O(t^9)\notag\\
&=(b_7+53)t^7\notag\\
&\qquad+\bigl[ b_8-4(a_n-a_{n-1}+2c_n-4c_{n-1}+c_{n-2})\notag\\
&\qquad\qquad-(b_7+53)(n-5)+36n-300\bigr]t^8-2t^{n+3}+\O(t^9)\notag\\
&=(b_7+53)t^7+\bigl[b_8-160-(b_7+53)(n-5)\bigr]t^8-2t^{n+3}+\O(t^9),\cr
\end{align}
while (\ref{BBBB}) becomes
\beq\label{BBBB2}
B_n(u)-\sum_{m=0}^\infty t^{2m}B_{n+m}(0)=(b_7+53)t^8+\O(t^9)
\eeq
giving
\beq\label{fn7}
t^{-2n}(f_{n+1}-f_n)=(b_7+53)t^7+\bigl[b_8-160-(b_7+53)(n-4)\bigr]t^8
-2t^{n+3}+\O(t^9).
\eeq        
Finally, if $b_7 = -53$ then
\begin{align}\label{order9A}
A_n&(u)-A_n(0)-t^2A_{n+1}(u)\notag\\
&=\bigl[4t^5-8t^6+36t^7+(b_8-258)t^8+(b_9+C)t^9\bigr]P_n \notag\\
&\qquad+\bigl(-t^3+3t^4-3t^5+19t^6+Ct^7\bigr)Q_n^1+(8t^5+Ct^6)Q_n^2\notag\\
&\qquad+\bigl(-2t^3+5t^4+Ct^5\bigr)R_n^1-(t^3+Ct^4)R_n^2-(4t^3+Ct^4)S_n+\O(t^9)\notag\\
&=(b_8-160)t^8+\bigl[ b_9+4(d_n-d_{n-1}) + 8a_n - 12a_{n-1} -4(a_{n-1}-a_{n-2})\notag\\
&\qquad\qquad - 24c_{n-1} -8c_{n-2} - 4c_{n-3} -92n + C\bigr]t^9-2t^{n+3}+\O(t^{10})\notag\\
&=(b_8-160)t^8+\bigl[b_9-C-(b_8-160)(n-5)\bigr]t^9-2t^{n+3}+\O(t^{10}),\cr
\end{align}
while from \eqref{BBB}---\eqref{B4}, \eqref{Beu}, \eqref{BBBB2},
\begin{align}\label{order9B}
B_n&(u)-\sum_{m=0}^\infty t^{2m}B_{n+m}(0)\notag\\
&=\bigl[t^4+2t^5+6t^6-2t^7+(b_8-96)t^8\bigr]\ti P_n+(11t^4-16t^5+69t^6)\ti Q_n\notag\\
&\qquad+(-t^3-6t^4+11t^5-28t^6)\ti Q_n^1-(9t^4-20t^5)\ti Q_n^2+(-2t^3+8t^4)\ti R_n^1\notag\\
&\qquad-t^3\ti R_n^2 +\O(t^{10})\notag \\
&= \bigl[ b_8+\ti d_n - \ti d_{n+2} -2\ti a_n +4\ti a_{n-1}-\ti c_{n+3} -4\ti c_{n+1} +6\ti c_n +20\ti c_{n-1} -4\ti c_{n-2}\notag\\
&\qquad +87n+130 \bigr]t^9 + \O(t^{10})\notag \\
&= (b_8-26)t^9 +\O(t^{10}),\cr
\end{align}
giving
\beq\label{fn8}
t^{-2n}(f_{n+1}-f_n)=(b_8-160)t^8+\bigl[b_9-C-(b_8-160)(n-4)\bigr]t^8
-2t^{n+3}+\O(t^9).
\eeq        

Now, collecting (\ref{fn3}), (\ref{fn4}), (\ref{fn5}), (\ref{fn6}),
(\ref{fn7}), (\ref{fn8}) gives: 

\bigskip\noindent{\bf Proposition 2:} 
The following are valid for sufficiently small $t$.
\begin{itemize}
\item
If $b_3>-1$, or $b_3=-1,\,b_4>4$, then 
\beq
f_{n+1}-f_n>0\,,\qquad n\ge2\,,
\eeq
and phases $3, 4, \dots$ are unstable relative to phase 2.
\item
If $b_3=-1,\,b_4=4$, and $-6<b_5<-4$, then 
\beqa
t^{-4}(f_3-f_2)&\simeq&(b_5+4)t^5<0\,,\cr
t^{-2n}(f_{n+1}-f_n)&\simeq&(b_5+6)t^5>0\,,\qquad n\ge3\,,
\eeqa
and phase $3$ is stable relative to phase 2 and to phases $4, 5, \dots$.
\item
If $b_3=-1,\,b_4=4,\,b_5=-6$, and $\472<b_6<{\frac{51}{2}}$, then
\beqa
t^{-4}(f_3-f_2)&\simeq&-2t^5<0\,,\cr
t^{-6}(f_4-f_3)&\simeq&(b_6-{\frac{51}{2}})t^{6}<0\,,\cr
t^{-2n}(f_{n+1}-f_n)&\simeq&(b_6-\472)t^{6}>0\,,\qquad n\ge4\,,
\eeqa
and phase $4$ is stable relative to phases 2, 3 and to phases $5, 6, \dots$.
\item
If $b_3=-1,\,b_4=4,\,b_5=-6,\,b_6=\472$, and $-53<b_7<-51$, then
\beqa
t^{-2n}(f_{n+1}-f_n)&\simeq&-2t^{n+3}<0\,,\qquad 2\le n\le3\,,\cr
t^{-8}(f_5-f_4)&\simeq&(b_7+51)t^{7}<0\,,\cr
t^{-2n}(f_{n+1}-f_n)&\simeq&(b_7+53)t^{7}>0\,,\qquad n\ge5\,,
\eeqa
and phase $5$ is stable relative to phases 2, 3, 4 and to phases $6, 7, \dots$.
\item
If $b_3=-1,\,b_4=4,\,b_5=-6,\,b_6=\472,\,b_7=-53$, and
$160<b_8<162$, then
\beqa
t^{-2n}(f_{n+1}-f_n)&\simeq&-2t^{n+3}<0\,,\qquad 2\le n\le4\,,\cr
t^{-10}(f_6-f_5)&\simeq&(b_8-162)t^{8}<0\,,\cr
t^{-2n}(f_{n+1}-f_n)&\simeq&(b_8-160)t^{8}>0\,,\qquad n\ge6\,,
\eeqa
and phase $6$ is stable relative to phases 2, 3, 4, 5 and to phases $7, 8, \dots$.
\item
There exists $B_9$ as follows.
If $b_3=-1,\,b_4=4,\,b_5=-6,\,b_6=\472,\,b_7=-53,b_8=160$ and $B_9 < b_9 < B_9+2$, then
\beqa
t^{-2n}(f_{n+1}-f_n)&\simeq&-2t^{n+3}<0\,,\qquad 2\le n\le5\,,\cr
t^{-10}(f_7-f_6)&\simeq&(b_9-B_9-2)t^{9}<0\,,\cr
t^{-2n}(f_{n+1}-f_n)&\simeq&(b_9-B_9)t^{9}>0\,,\qquad n\ge7\,,
\eeqa
and phase 7 is stable relative to phases $2,3,4,5,6$ and to phases $8,9,\dots$
\end{itemize}

\newpage
\section{Phases 0, 1, 2}\label{012}
For $n=0$, (\ref{Zn}) takes the form
\beq\label{Zn0}
Z_0^\La=e^{u|\La_1|}\sum_{\{\ga\}}\prod_\ga \psi(\ga)
\eeq
with
\beq\label{psi}
\psi(\ga)=t^{{\frac{1}{2}}|\ga|-|\ga\cap I_n|}
e^{-u|\ga\cap\{z={\frac{1}{2}}\}|}
\eeq
so that
\beqa
\log(Z_0^\La)&=&u|\La_1|+\sum_\om\psi^T(\om)\cr
&=&u|\La_1|+\sum_{\om\in W\atop\om\approx W_N}\psi^T(\om)
+\sum_{\om\approx W,W_N}\vphi_2^T(\om)
\eeqa
while
\beqa
\log(Z_1^\La)&=&\sum_{\om\in I_1\atop\om\approx W_N}\vphi^T(\om)
+\sum_{\om\sim I_1,\om\in W\atop \om\approx W_N}\vphi^T(\om)
+\sum_{\om\sim I_1\atop \om\approx W,W_N}\vphi^T(\om)\cr
&=&\sum_{\om\in I_1\atop\om\approx W_N}\vphi^T(\om)
+\sum_{\om\sim I_1,\om\in W\atop \om\approx W_N}\vphi_1^T(\om)
+\sum_{\om\sim I_1\atop \om\approx W,W_N}\vphi_2^T(\om)
\eeqa
Therefore
\begin{align}
\log&(Z_0^\La/Z_1^\La)\\
&=u|\La_1|+\sum_{\om\in W\atop\om\approx W_N}\psi^T(\om)
-\sum_{\om\in I_1\atop\om\approx W_N}\vphi^T(\om)
+\sum_{\om\approx W,W_N\atop\om\not\sim I_1}\vphi_2^T(\om)
-\sum_{\om\approx W_N,\om\in W\atop\om\sim I_1}\vphi_1^T(\om),\notag
\end{align}
giving
\beqa\label{f10}
f_1-f_0&=&u+(e^{-u}t^2+2e^{-2u}t^3)-(t^2+e^ut^2+2t^3+2e^{2u}t^3)+t^3+\O(t^4)
\cr
&=&(b_3-1)t^3+\O(t^4)
\eeqa

For $n=1$, in order to use (\ref{AB}), we need 
$A_1(u), A_2(u), B_1(u), B_2(0)$.  The expansion
\beq
t^2A_1(u)=e^ut^2+2e^{2u}t^3+6e^{3u}t^4+e^{4u}t^4-e^ut^4-\12e^{2u}t^4-2e^{2u}t^4
+\O(t^5)
\eeq
gives
\begin{align}
t^2\bigl(A_1(u)-A_1(0)\bigr)&=(e^u-1)t^2+2(e^{2u}-1)t^3+6(e^{3u}-1)t^4
+(e^{4u}-1)t^4\notag\\
&\qquad-(e^u-1)t^4-\12(e^{2u}-1)t^4-2(e^{2u}-1)t^4+\O(t^7)\notag\\
&=t^4+(b_3+4)t^5+(b_4+4b_3+17)t^6+\O(t^7).
\end{align}
We then compute $A_2(u)$ using $P_2, Q_2$:
\beqa
P_2&=&(1+4t-t-4t^2)+(12t^2+6t^2+4t^2-6t^2)+\O(t^3)\cr
&=&1+3t+12t^2+\O(t^3)
\eeqa
where the first parenthesis is adapted from Fig. 2 and the second from Fig. 3.
Also, adapted from Fig. 7,
\beqa
Q_2&=&4t^2-4t^3+12t^3-5t^3+2t^2+12t^3+\O(t^4)\cr
&=&6t^2+15t^3+\O(t^4)
\eeqa
giving
\beqa
t^4A_2(u)&=&e^ut^4(1+3t+12t^2)+6e^{2u}t^6+\O(t^7)\cr
&=&t^4+3t^5+19t^6+\O(t^7)
\eeqa
and
\beq
t^2A_1(u)-t^2A_1(0)-t^4A_2(u)=(b_3+1)t^5+(b_4+4b_3-2)t^6+\O(t^7).
\eeq
Then
\beq
t^2B_1(u)=e^ut^3+2e^{2u}t^5-e^ut^5+\O(t^7),
\eeq
\beq
t^4B_2(0)=t^5+\O(t^7),
\eeq
\beq
t^2B_1(u)-t^2B_1(0)-t^4B_2(0)=b_3t^6+\O(t^7)
\eeq
so that finally
\beq\label{f21}
f_2-f_1=(b_3+1)t^5+(b_4+3b_3-2)t^6+\O(t^7),
\eeq
which completes the derivation of (\ref{123456}).

\newpage
\section{Recursion diagrams, $n\ge3$}\label{RecDia}
For the recursion relations \eqref{induc} relating
$n$ to $n+1$, we consider ways in which a cluster 
$\om\in I_n,W$ can be extended to produce a new 
$\om' \in I_{n+1},W$.  
One choice is that one or more polymers in $\om$ may be extended without adding polymers 
or changing incompatibility relations within $\om$. Then the
combinatoric factor in (\ref{phiT}) is unchanged, only the $\vphi(\ga)$ for the 
extended polymers change, and it remains to find a geometric factor, the
number of ways to extend the polymer, or the number of diagrams of a
given type. 

Next, one may have $\om'=\om\cup\{\ga'\}$ with the new polymer
incompatible with only one polymer from $\om$. Then (\ref{phiT}) gives
$\vphi^T(\om')=-\vphi(\ga')\vphi^T(\om)$, with $\vphi^T(\om)$ taking
into account possible polymer extensions as in the first case.   

Next, one may have $\om'=\om\cup\{\ga'\}$ with the new polymer
incompatible with two polymers $\ga_1,\ga_2$ from $\om$. 
At the order considered here, one may assume that $\ga_1\not\sim\ga_2$
and that $\om=\{\ga_1,\ga_2\}$ or $\om=\{\ga_0,\ga_1,\ga_2\}$. 
Then (\ref{phiT}) gives 
\beq\label{m2}
\vphi^T(\om')=-2\vphi(\ga')\vphi^T(\om)\,,
\eeq
with $\vphi^T(\om)$ taking
into account possible polymer extensions as in the first
case. Formula (\ref{m2}) occurs in the 2nd and 3rd diagrams in the 2nd line
for $Q_{n+1}^1$, and in the 3rd and 5th diagrams in the 2nd line for
$Q_{n+1}^2$.    
 

\vskip .8 cm

\setlength{\unitlength}{1900sp}%
\begin{picture}(13254,2274)(259,-1423)
\thinlines
{\put(451,839){\line( 1, 0){450}}
\put(901,839){\line( 1,-6){225}}
\put(1126,-511){\line( 0,-1){855}}
\put(1126,-1366){\line( 1, 0){900}}
\put(2026,-1366){\line( 0, 1){405}}
\put(2026,-961){\line(-1, 0){450}}
\put(1576,-961){\line( 0, 1){450}}
\put(1576,-511){\line( 1, 6){225}}
\put(1801,839){\line( 1, 0){450}}
}%
{\put(2701,839){\line( 1, 0){450}}
\put(3151,839){\line( 1,-6){225}}
\put(3376,-511){\line( 0,-1){450}}
\put(3376,-961){\line( 1, 0){450}}
\put(3826,-961){\line( 0, 1){450}}
\put(3826,-511){\line( 1, 6){225}}
\put(4051,839){\line( 1, 0){450}}
}%
{\put(3376,-1366){\framebox(900,360){}}
}%
{\put(4951,839){\line( 1, 0){450}}
\put(5401,839){\line(-1,-6){225}}
\put(5176,-511){\line( 0,-1){855}}
\put(5176,-1366){\line( 1, 0){900}}
\put(6076,-1366){\line( 0, 1){855}}
\put(6076,-511){\line(-1, 6){225}}
\put(5851,839){\line( 1, 0){450}}
}%
{\multiput(271,-1411)(114.54545,0.00000){116}{\line( 1, 0){ 57.273}}
}%
{\put(10801,839){\line( 1, 0){450}}
\put(11251,839){\line( 1,-6){225}}
\put(11476,-511){\line( 0,-1){450}}
\put(11476,-961){\line( 1, 0){450}}
\put(11926,-961){\line( 0, 1){450}}
\put(11926,-511){\line( 1, 6){225}}
\put(12151,839){\line( 1, 0){450}}
}%
{\put(11971,-511){\line( 0,-1){855}}
\put(11971,-1366){\line( 1, 0){855}}
\put(12826,-1366){\line( 0, 1){405}}
\put(12826,-961){\line(-1, 0){450}}
\put(12376,-961){\line( 0, 1){450}}
\put(12376,-511){\line(-1, 0){405}}
\put(11971,-511){\line( 0, 1){  0}}
}%
{\put(9001,839){\line( 1, 0){495}}
\put(9496,839){\line(-1,-5){270}}
\put(9226,-511){\line( 0,-1){450}}
\put(9226,-961){\line( 1, 0){900}}
\put(10126,-961){\line( 0, 1){450}}
\put(10126,-511){\line(-1, 6){225}}
\put(9901,839){\line( 1, 0){450}}
}%
{\put(9676,-1366){\framebox(900,360){}}
}%
{\put(6976,839){\line( 1, 0){450}}
\put(7426,839){\line(-1,-6){225}}
\put(7201,-511){\line( 0,-1){450}}
\put(7201,-961){\line( 1, 0){450}}
\put(7651,-961){\line( 0,-1){405}}
\put(7651,-1366){\line( 1, 0){900}}
\put(8551,-1366){\line( 0, 1){405}}
\put(8551,-961){\line(-1, 0){450}}
\put(8101,-961){\line( 0, 1){450}}
\put(8101,-511){\line(-1, 6){225}}
\put(7876,839){\line( 1, 0){450}}
}%
\put(00,-2200){\footnotesize $Q_{n+1}^1=$}
\put(1300,-2200){\footnotesize $4t^2P_n$}
\put(3000,-2200){\footnotesize $-4t^3P_n$}
\put(5100,-2200){\footnotesize $+tQ_n^1$}
\put(7200,-2200){\footnotesize $+6t^2Q_n^1$}
\put(9100,-2200){\footnotesize $-7t^3Q_n^1$}
\put(11300,-2200){\footnotesize $+8t^2Q_n^2$}
\end{picture}%

\vskip2cm
\setlength{\unitlength}{1780sp}%
\begin{picture}(14199,2274)(-11,-1423)
\thinlines
{\put(  1,839){\line( 1, 0){450}}
\put(451,839){\line( 1,-6){225}}
\put(676,-511){\line( 0,-1){450}}
\put(676,-961){\line( 1, 0){450}}
\put(1126,-961){\line( 0, 1){450}}
\put(1126,-511){\line( 1, 6){225}}
\put(1351,839){\line( 1, 0){450}}
}%
{\put(1171,-961){\framebox(405,450){}}
}%
{\put(1171,-1366){\framebox(855,360){}}
}%
{\put(2251,839){\line( 1, 0){450}}
\put(2701,839){\line( 1,-6){225}}
\put(2926,-511){\line( 0,-1){450}}
\put(2926,-961){\line( 1, 0){450}}
\put(3376,-961){\line( 0, 1){450}}
\put(3376,-511){\line( 1, 6){225}}
\put(3601,839){\line( 1, 0){450}}
}%
{\put(3421,-961){\framebox(405,450){}}
}%
{\put(2926,-1366){\framebox(900,360){}}
}%
{\put(4501,839){\line( 1, 0){450}}
\put(4951,839){\line( 1,-6){225}}
\put(5176,-511){\line( 0,-1){450}}
\put(5176,-961){\line( 1, 0){450}}
\put(5626,-961){\line( 0, 1){450}}
\put(5626,-511){\line( 1, 6){225}}
\put(5851,839){\line( 1, 0){450}}
}%
\thicklines
{\put(5221,-916){\dashbox{57}(450,450){}}
}%
\thinlines
{\put(5176,-1366){\framebox(900,360){}}
}%
{\put(6751,839){\line( 1, 0){450}}
\put(7201,839){\line(-1,-3){450}}
\put(6751,-511){\line( 0,-1){855}}
\put(6751,-1366){\line( 1, 0){900}}
\put(7651,-1366){\line( 0, 1){405}}
\put(7651,-961){\line( 1, 0){450}}
\put(8101,-961){\line( 0, 1){450}}
\put(8101,-511){\line(-1, 3){450}}
\put(7651,839){\line( 1, 0){450}}
}%
{\put(8551,839){\line( 1, 0){450}}
\put(9001,839){\line(-1,-3){450}}
\put(8551,-511){\line( 0,-1){450}}
\put(8551,-961){\line( 1, 0){900}}
\put(9451,-961){\line( 0,-1){405}}
\put(9451,-1366){\line( 1, 0){900}}
\put(10351,-1366){\line( 0, 1){405}}
\put(10351,-961){\line(-1, 0){450}}
\put(9901,-961){\line( 0, 1){450}}
\put(9901,-511){\line(-1, 3){450}}
\put(9451,839){\line( 1, 0){450}}
}%
{\put(12601,839){\line( 1, 0){450}}
\put(13051,839){\line(-1,-6){225}}
\put(12826,-511){\line( 0,-1){855}}
\put(12826,-1366){\line( 1, 0){900}}
\put(13726,-1366){\line( 0, 1){855}}
\put(13726,-511){\line(-1, 6){225}}
\put(13501,839){\line( 1, 0){450}}
}%
{\put(13771,-961){\framebox(405,450){}}
}%
{\multiput(271,-1411)(114.44444,0.00000){122}{\line( 1, 0){ 57.222}}
}%
{\put(10801,839){\line( 1, 0){450}}
\put(11251,839){\line(-1,-3){450}}
\put(10801,-511){\line( 0,-1){855}}
\put(10801,-1366){\line( 1, 0){450}}
\put(11251,-1366){\line( 0, 1){405}}
\put(11251,-961){\line( 1, 0){450}}
\put(11701,-961){\line( 0,-1){405}}
\put(11701,-1366){\line( 1, 0){450}}
\put(12151,-1366){\line( 0, 1){855}}
\put(12151,-511){\line(-1, 3){450}}
\put(11701,839){\line( 1, 0){450}}
}%

\put(400,-2200){\footnotesize $-6t^3Q_n^{2a}$}
\put(2800,-2200){\footnotesize $-2t^3Q_n^{2a}$}
\put(4900,-2200){\footnotesize $-8t^3Q_n^{2b}$}
\put(6900,-2200){\footnotesize $+2tR_n^1$}
\put(8800,-2200){\footnotesize $+8t^2R_n$}
\put(11000,-2200){\footnotesize $+t^2R_n$}
\put(12800,-2200){\footnotesize $+tR_n^2$}
\put(13800,-2200){\footnotesize $+\dots$}
\end{picture}%

\vskip 1 cm
\begin{center}
\begin{minipage}{9cm}
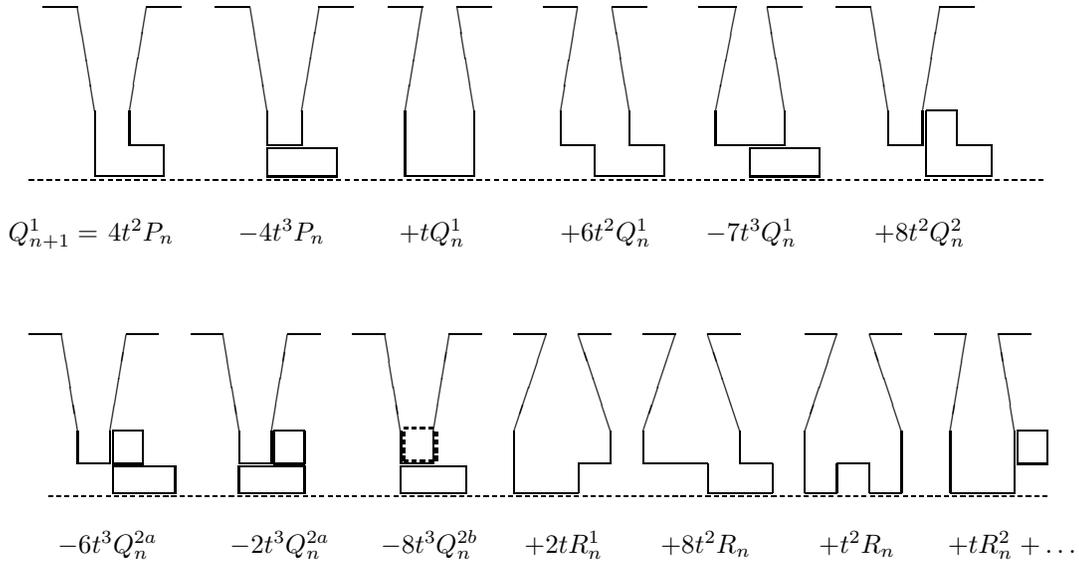
{\footnotesize{\centerline{Fig. 10. Recursion for $Q_{n+1}^1$.}}} 
\end{minipage}
\end{center}
\vskip .8 cm

\setlength{\unitlength}{1900sp}%
\begin{picture}(12129,2274)(259,-1423)
\thinlines
{\put(10801,839){\line( 1, 0){450}}
\put(11251,839){\line(-1,-6){225}}
\put(11026,-511){\line( 0,-1){450}}
\put(11026,-961){\line( 1, 0){900}}
\put(11926,-961){\line( 0, 1){450}}
\put(11926,-511){\line(-1, 6){225}}
\put(11701,839){\line( 1, 0){450}}
}%
{\put(11026,-1366){\framebox(450,315){}}
}%
\thicklines
{\put(11071,-1321){\dashbox{57}(450,315){}}
}%
\thinlines
{\multiput(271,-1411)(114.73934,0.00000){106}{\line( 1, 0){ 57.370}}
}%
{\put(9001,839){\line( 1, 0){450}}
\put(9451,839){\line(-1,-6){225}}
\put(9226,-511){\line( 0,-1){450}}
\put(9226,-961){\line( 1, 0){900}}
\put(10126,-961){\line( 0, 1){450}}
\put(10126,-511){\line(-1, 6){225}}
\put(9901,839){\line( 1, 0){450}}
}%
{\put(9721,-1366){\framebox(405,360){}}
}%
{\put(9226,-1366){\framebox(450,360){}}
}%
{\put(7201,839){\line( 1, 0){450}}
\put(7651,839){\line(-1,-6){225}}
\put(7426,-511){\line( 0,-1){450}}
\put(7426,-961){\line( 1, 0){900}}
\put(8326,-961){\line( 0, 1){450}}
\put(8326,-511){\line(-1, 6){225}}
\put(8101,839){\line( 1, 0){450}}
}%
{\put(7426,-1366){\framebox(450,360){}}
}%
{\put(6931,-1366){\framebox(450,360){}}
}%
{\put(5401,839){\line( 1, 0){450}}
\put(5851,839){\line(-1,-6){225}}
\put(5626,-511){\line( 0,-1){855}}
\put(5626,-1366){\line( 1, 0){450}}
\put(6076,-1366){\line( 0, 1){405}}
\put(6076,-961){\line( 1, 0){450}}
\put(6526,-961){\line( 0, 1){450}}
\put(6526,-511){\line(-1, 6){225}}
\put(6301,839){\line( 1, 0){450}}
}%
{\put(5131,-1366){\framebox(450,360){}}
}%
{\put(2926,839){\line( 1, 0){450}}
\put(3376,839){\line( 1,-6){225}}
\put(3601,-511){\line( 0,-1){450}}
\put(3601,-961){\line( 1, 0){450}}
\put(4051,-961){\line( 0, 1){450}}
\put(4051,-511){\line( 1, 6){225}}
\put(4276,839){\line( 1, 0){450}}
}%
{\put(3601,-1366){\framebox(450,360){}}
}%
{\put(3106,-1366){\framebox(450,360){}}
}%
{\put(1846,-1366){\framebox(405,450){}}
}%
{\put(676,839){\line( 1, 0){450}}
\put(1126,839){\line( 1,-6){225}}
\put(1351,-511){\line( 0,-1){855}}
\put(1351,-1366){\line( 1, 0){450}}
\put(1801,-1366){\line( 0, 1){855}}
\put(1801,-511){\line( 1, 6){225}}
\put(2026,839){\line( 1, 0){450}}
}%
\put(-400,-2200){\footnotesize $Q_{n+1}^2=$}
\put(1100,-2200){\footnotesize $-5t^3P_n$}
\put(3000,-2200){\footnotesize $+5t^4P_n$}
\put(5200,-2200){\footnotesize $-10t^3Q_n^1$}
\put(7200,-2200){\footnotesize $+6t^4Q_n^1$}
\put(9000,-2200){\footnotesize $+2t^4Q_n^1$}
\put(11000,-2200){\footnotesize $+2t^4Q_n^1$}
\end{picture}%

\vskip2cm


\begin{picture}(11454,2274)(259,-1423)
\thinlines
{\multiput(271,-1411)(113.73134,0.00000){101}{\line( 1, 0){ 56.866}}
}%
{\put(9901,839){\line( 1, 0){450}}
\put(10351,839){\line( 1,-6){225}}
\put(10576,-511){\line( 0,-1){855}}
\put(10576,-1366){\line( 1, 0){450}}
\put(11026,-1366){\line( 0, 1){855}}
\put(11026,-511){\line( 1, 6){225}}
\put(11251,839){\line( 1, 0){450}}
}%
\thicklines
{\put(10531,-916){\dashbox{57}(450,405){}}
}%
{\put(10621,-1366){\dashbox{57}(450,405){}}
}%
\thinlines
{\put(7651,839){\line( 1, 0){450}}
\put(8101,839){\line( 1,-6){225}}
\put(8326,-511){\line( 0,-1){855}}
\put(8326,-1366){\line( 1, 0){450}}
\put(8776,-1366){\line( 0, 1){855}}
\put(8776,-511){\line( 1, 6){225}}
\put(9001,839){\line( 1, 0){450}}
}%
\thicklines
{\put(8281,-961){\dashbox{57}(450,450){}}
}%
\thinlines
{\put(8821,-1366){\framebox(405,405){}}
}%
{\put(5401,839){\line( 1, 0){450}}
\put(5851,839){\line( 1,-6){225}}
\put(6076,-511){\line( 0,-1){855}}
\put(6076,-1366){\line( 1, 0){450}}
\put(6526,-1366){\line( 0, 1){855}}
\put(6526,-511){\line( 1, 6){225}}
\put(6751,839){\line( 1, 0){450}}
}%
{\put(6571,-916){\framebox(405,405){}}
}%
{\put(6571,-1366){\framebox(405,405){}}
}%
{\put(3151,839){\line( 1, 0){450}}
\put(3601,839){\line( 1,-6){225}}
\put(3826,-511){\line( 0,-1){855}}
\put(3826,-1366){\line( 1, 0){450}}
\put(4276,-1366){\line( 0, 1){855}}
\put(4276,-511){\line( 1, 6){225}}
\put(4501,839){\line( 1, 0){450}}
}%
{\put(3376,-1366){\framebox(405,405){}}
}%
{\put(4321,-916){\framebox(405,405){}}
}%
{\put(901,839){\line( 1, 0){450}}
\put(1351,839){\line( 1,-6){225}}
\put(1576,-511){\line( 0,-1){855}}
\put(1576,-1366){\line( 1, 0){450}}
\put(2026,-1366){\line( 0, 1){855}}
\put(2026,-511){\line( 1, 6){225}}
\put(2251,839){\line( 1, 0){450}}
}%
{\put(2071,-1366){\framebox(405,855){}}
}%
\put(1100,-2200){\footnotesize $+t^2Q_n^2$}
\put(3200,-2200){\footnotesize $-8t^3Q_n^{2a}$}
\put(5700,-2200){\footnotesize $-4t^3Q_n^{2a}$}
\put(8200,-2200){\footnotesize $-8t^3Q_n^{2b}$}
\put(10200,-2200){\footnotesize $-4t^3Q_n^{2b}$}
\put(12000,-2200){\footnotesize $+\dots$}
\end{picture}%
\vskip 1cm
\begin{center}
\begin{minipage}{9cm}
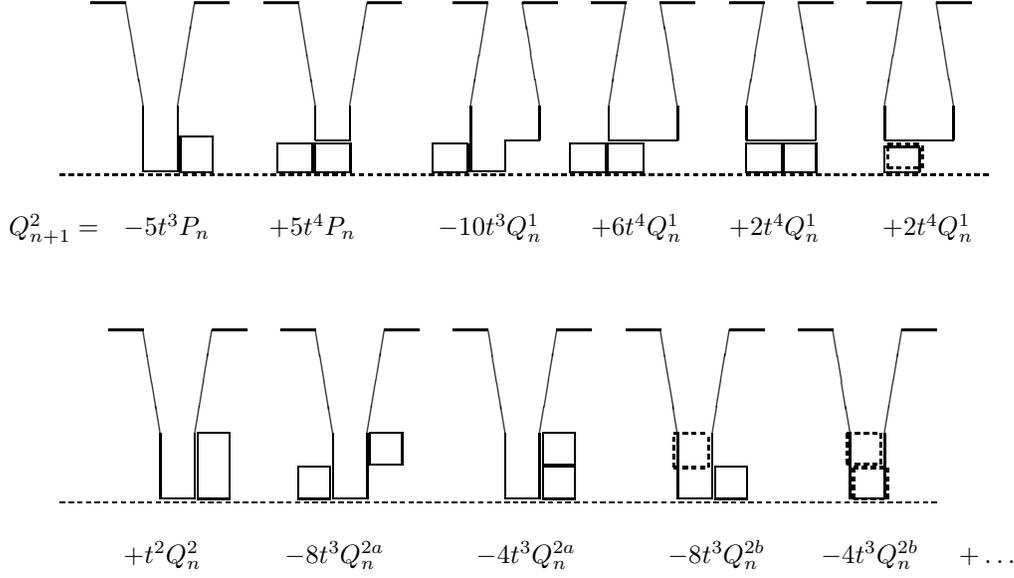
{\footnotesize{\centerline{Fig. 11. Recursion for $Q_{n+1}^2$.}}} 
\end{minipage}
\end{center}
\vskip .8 cm 

Next, one may have $\om'=\om\cup\{\ga'_1,\ga'_2\}$ with each of
$\ga'_1,\ga'_2$ incompatible with at most one polymer in $\om$.
If $\ga'_2\not\sim\ga'_1\not\sim\om$ and $\ga'_2\sim\om$, or 
$\ga'_2\not\sim\om\not\sim\ga'_1$ and $\ga'_2\sim\ga'_1$,
then (\ref{phiT}) gives
$\vphi^T(\om')=\vphi(\ga'_1)\vphi(\ga'_2)\vphi^T(\om)$, with $\vphi^T(\om)$ 
taking into account possible polymer extensions as in the first case.   
If $\ga'_2\not\sim\ga'_1\not\sim\om\not\sim\ga'_2$ and $\ga'_2\neq\ga'_1$, 
and $\ga'_1$ and $\ga'_2$ are incompatible with the same polymer in
$\om$, then (\ref{phiT}) gives 
\beq\label{m22}
\vphi^T(\om')=2\vphi(\ga'_1)\vphi(\ga'_2)\vphi^T(\om)\,,
\eeq
with $\vphi^T(\om)$ taking into account possible polymer extensions as
in the first case. Formula (\ref{m22}) occurs in the 5th diagram
in the 1st line for $Q_{n+1}^2$ and in the 5th diagram for $R_{n+1}^2$   
and in the last diagram for $R_{n+1}^3$. If $\ga'_2=\ga'_1$, then 
$\vphi^T(\om')=\vphi(\ga'_1)\vphi(\ga'_2)\vphi^T(\om)$, with $\vphi^T(\om)$ 
taking into account possible polymer extensions as in the first case.   
 

\bigskip
\newpage

\begin{picture}(13254,2274)(259,-1423)
\thinlines
{\put(2701,839){\line( 1, 0){450}}
\put(3151,839){\line( 1,-6){225}}
\put(3376,-511){\line( 0,-1){450}}
\put(3376,-961){\line( 1, 0){450}}
\put(3826,-961){\line( 0, 1){450}}
\put(3826,-511){\line( 1, 6){225}}
\put(4051,839){\line( 1, 0){450}}
}%
{\multiput(271,-1411)(100,0.00000){116}{\line( 1, 0){ 57.273}}
}%
{\put(451,839){\line( 1, 0){450}}
\put(901,839){\line( 1,-6){225}}
\put(1126,-511){\line( 0,-1){450}}
\put(1126,-961){\line(-1, 0){450}}
\put(676,-961){\line( 0,-1){405}}
\put(676,-1366){\line( 1, 0){1350}}
\put(2026,-1366){\line( 0, 1){405}}
\put(2026,-961){\line(-1, 0){450}}
\put(1576,-961){\line( 0, 1){450}}
\put(1576,-511){\line( 1, 6){225}}
\put(1801,839){\line( 1, 0){450}}
}%
{\put(2926,-1366){\framebox(1350,360){}}
}%
{\put(4951,839){\line( 1, 0){450}}
\put(5401,839){\line(-1,-6){225}}
\put(5176,-511){\line( 0,-1){450}}
\put(5176,-961){\line( 0,-1){405}}
\put(5176,-1366){\line( 1, 0){450}}
\put(5626,-1366){\line( 1, 0){900}}
\put(6526,-1366){\line( 0, 1){405}}
\put(6526,-961){\line(-1, 0){450}}
\put(6076,-961){\line( 0, 1){450}}
\put(6076,-511){\line(-1, 6){225}}
\put(5851,839){\line( 1, 0){450}}
}%
{\put(9901,839){\line( 1, 0){450}}
\put(10351,839){\line(-1,-3){450}}
\put(9901,-511){\line( 0,-1){855}}
\put(9901,-1366){\line( 1, 0){1350}}
\put(11251,-1366){\line( 0, 1){855}}
\put(11251,-511){\line(-1, 3){450}}
\put(10801,839){\line( 1, 0){450}}
}%
{\put(7201,839){\line( 1, 0){450}}
\put(7651,839){\line(-1,-6){225}}
\put(7426,-511){\line( 0,-1){450}}
\put(7426,-961){\line( 1, 0){450}}
\put(7876,-961){\line( 0,-1){405}}
\put(7876,-1366){\line( 1, 0){1350}}
\put(9226,-1366){\line( 0, 1){405}}
\put(9226,-961){\line(-1, 0){900}}
\put(8326,-961){\line( 0, 1){450}}
\put(8326,-511){\line(-1, 6){225}}
\put(8101,839){\line( 1, 0){450}}
}%
\put(-400,-2200){\footnotesize $R_{n+1}^1=$}
\put(1100,-2200){\footnotesize $18t^4P_n$}
\put(3000,-2200){\footnotesize $-18t^5P_n$}
\put(5200,-2200){\footnotesize $+6t^3Q_n^1$}
\put(7500,-2200){\footnotesize $+24t^4Q_n^1$}
\put(10000,-2200){\footnotesize $+t^2R_n$}
\put(11800,-2200){\footnotesize $+\dots$}
\end{picture}%

\vskip2cm

\begin{picture}(13254,2274)(259,-1423)
\thinlines
{\multiput(271,-1411)(114.54545,0.00000){116}{\line( 1, 0){ 57.273}}
}%
{\put(451,839){\line( 1, 0){450}}
\put(901,839){\line( 1,-6){225}}
\put(1126,-511){\line( 0,-1){450}}
\put(1126,-961){\line( 0, 1){  0}}
\put(1126,-961){\line( 0,-1){405}}
\put(1126,-1366){\line( 1, 0){900}}
\put(2026,-1366){\line( 0, 1){405}}
\put(2026,-961){\line(-1, 0){450}}
\put(1576,-961){\line( 0, 1){450}}
\put(1576,-511){\line( 1, 6){225}}
\put(1801,839){\line( 1, 0){450}}
}%
{\put(676,-1366){\framebox(405,405){}}
}%
{\put(2701,839){\line( 1, 0){450}}
\put(3151,839){\line( 1,-6){225}}
\put(3376,-511){\line( 0,-1){855}}
\put(3376,-1366){\line( 1, 0){450}}
\put(3826,-1366){\line( 0, 1){855}}
\put(3826,-511){\line( 1, 6){225}}
\put(4051,839){\line( 1, 0){450}}
}%
{\put(3871,-1366){\framebox(855,405){}}
}%
{\put(5626,-1366){\framebox(900,360){}}
}%
{\put(5176,-1366){\framebox(405,360){}}
}%
{\put(7876,-1366){\framebox(450,360){}}
}%
{\put(8371,-1366){\framebox(855,360){}}
}%
{\put(10126,-1366){\framebox(900,360){}}
}%
\thicklines
{\put(10081,-1366){\dashbox{57}(495,315){}}
}%
\thinlines
{\put(12871,-1366){\framebox(405,405){}}
}%
{\put(11701,839){\line( 1, 0){450}}
\put(12151,839){\line(-1,-6){225}}
\put(11926,-511){\line( 0,-1){855}}
\put(11926,-1366){\line( 1, 0){900}}
\put(12826,-1366){\line( 0, 1){855}}
\put(12826,-511){\line(-1, 6){225}}
\put(12601,839){\line( 1, 0){450}}
}%
{\put(9451,839){\line( 1, 0){450}}
\put(9901,839){\line( 1,-6){225}}
\put(10126,-511){\line( 0,-1){450}}
\put(10126,-961){\line( 1, 0){450}}
\put(10576,-961){\line( 0, 1){450}}
\put(10576,-511){\line( 1, 6){225}}
\put(10801,839){\line( 1, 0){450}}
}%
{\put(7201,839){\line( 1, 0){450}}
\put(7651,839){\line( 1,-6){225}}
\put(7876,-511){\line( 0,-1){450}}
\put(7876,-961){\line( 1, 0){450}}
\put(8326,-961){\line( 0, 1){450}}
\put(8326,-511){\line( 1, 6){225}}
\put(8551,839){\line( 1, 0){450}}
}%
{\put(4951,839){\line( 1, 0){450}}
\put(5401,839){\line( 1,-6){225}}
\put(5626,-511){\line( 0,-1){450}}
\put(5626,-961){\line( 1, 0){450}}
\put(6076,-961){\line( 0, 1){450}}
\put(6076,-511){\line( 1, 6){225}}
\put(6301,839){\line( 1, 0){450}}
}%
\put(-400,-2200){\footnotesize $R_{n+1}^2=$}
\put(900,-2200){\footnotesize $-32t^5P_n$}
\put(3000,-2200){\footnotesize $-16t^5P_n$}
\put(5200,-2200){\footnotesize $+28t^6P_n$}
\put(7500,-2200){\footnotesize $+12t^6P_n$}
\put(9800,-2200){\footnotesize $+8t^6P_n$}
\put(11800,-2200){\footnotesize $-8t^4Q_n^1$}
\put(13100,-2200){\footnotesize $+\dots$}
\end{picture}%

\vskip2cm
\begin{picture}(13254,2274)(259,-1423)
\thinlines
{\multiput(271,-1411)(114.54545,0.00000){116}{\line( 1, 0){ 57.273}}
}%
{\put(451,839){\line( 1, 0){450}}
\put(901,839){\line( 1,-6){225}}
\put(1126,-511){\line( 0,-1){855}}
\put(1126,-1366){\line( 1, 0){450}}
\put(1576,-1366){\line( 0, 1){855}}
\put(1576,-511){\line( 1, 6){225}}
\put(1801,839){\line( 1, 0){450}}
}%
{\put(1621,-1366){\framebox(405,405){}}
}%
{\put(2071,-1366){\framebox(405,405){}}
}%
{\put(2701,839){\line( 1, 0){450}}
\put(3151,839){\line( 1,-6){225}}
\put(3376,-511){\line( 0,-1){855}}
\put(3376,-1366){\line( 1, 0){450}}
\put(3826,-1366){\line( 0, 1){855}}
\put(3826,-511){\line( 1, 6){225}}
\put(4051,839){\line( 1, 0){450}}
}%
{\put(3871,-1366){\framebox(405,405){}}
}%
{\put(2926,-1366){\framebox(405,405){}}
}%
{\put(4951,839){\line( 1, 0){450}}
\put(5401,839){\line( 1,-6){225}}
\put(5626,-511){\line( 0,-1){855}}
\put(5626,-1366){\line( 1, 0){450}}
\put(6076,-1366){\line( 0, 1){855}}
\put(6076,-511){\line( 1, 6){225}}
\put(6301,839){\line( 1, 0){450}}
}%
{\put(6121,-1366){\framebox(405,405){}}
}%
\thicklines
{\put(6166,-1321){\dashbox{57}(405,405){}}
}%
\thinlines
{\put(7201,839){\line( 1, 0){450}}
\put(7651,839){\line( 1,-6){225}}
\put(7876,-511){\line( 0,-1){855}}
\put(7876,-1366){\line( 1, 0){450}}
\put(8326,-1366){\line( 0, 1){855}}
\put(8326,-511){\line( 1, 6){225}}
\put(8551,839){\line( 1, 0){450}}
}%
{\put(8371,-1366){\framebox(405,405){}}
}%
\thicklines
{\put(7831,-1321){\dashbox{57}(450,360){}}
}%
\put(-400,-2200){\footnotesize $R_{n+1}^3=$}
\put(900,-2200){\footnotesize $12t^6P_n$}
\put(3000,-2200){\footnotesize $+6t^6P_n$}
\put(5200,-2200){\footnotesize $+5t^6P_n$}
\put(7500,-2200){\footnotesize $+8t^6P_n$}
\put(10100,-2200){\footnotesize $+\dots$}
\end{picture}%

\vskip2cm
\setlength{\unitlength}{2500sp}%
\begin{picture}(9924,924)(439,-73)
\thinlines
{\put(451,389){\framebox(450,450){}}
}%
{\put(901,389){\framebox(450,450){}}
}%
{\put(4051,-61){\framebox(450,450){}}
}%
{\put(1801,-61){\framebox(450,450){}}
}%
{\put(2251,-61){\framebox(450,450){}}
}%
{\put(2701,-61){\framebox(450,450){}}
}%
{\put(3151,-61){\framebox(450,450){}}
}%
{\put(4501,-61){\framebox(450,450){}}
}%
{\put(4501,389){\framebox(450,450){}}
}%
{\put(4951,389){\framebox(450,450){}}
}%
{\put(6301,-61){\framebox(450,450){}}
}%
{\put(5851,-61){\framebox(450,450){}}
}%
{\put(6751,-61){\framebox(450,450){}}
}%
{\put(6751,389){\framebox(450,450){}}
}%
{\put(8101,-61){\framebox(450,450){}}
}%
{\put(7651,-61){\framebox(450,450){}}
}%
{\put(8101,389){\framebox(450,450){}}
}%
{\put(8551,-61){\framebox(450,450){}}
}%
{\put(9451,-61){\framebox(450,450){}}
}%
{\put(9901,-61){\framebox(450,450){}}
}%
{\put(9901,389){\framebox(450,450){}}
}%
{\put(9451,389){\framebox(450,450){}}
}%
{\put(901,-61){\framebox(450,450){}}
}%
{\put(451,-61){\framebox(450,450){}}
}%
{\put(451,-61){\line( 1, 1){450}}
}%
{\put(451,389){\line( 1,-1){450}}
}%
{\put(1801,-61){\line( 1, 1){450}}
}%
{\put(4051,-61){\line( 1, 1){450}}
}%
{\put(6301,-61){\line( 1, 1){450}}
}%
{\put(8101,-61){\line( 1, 1){450}}
}%
{\put(9451,-61){\line( 1, 1){450}}
}%
{\put(9901,-61){\line( 1, 1){450}}
}%
{\put(1801,389){\line( 1,-1){450}}
}%
{\put(4051,389){\line( 1,-1){450}}
}%
{\put(6301,389){\line( 1,-1){450}}
}%
{\put(8101,389){\line( 1,-1){450}}
}%
{\put(9451,389){\line( 1,-1){450}}
}%
{\put(9901,389){\line( 1,-1){450}}
}%
\put(-400,-800){\footnotesize $S_{n+1}=$}
\put(500,-800){\footnotesize $4t^5P_n$}
\put(2000,-800){\footnotesize $+12t^6P_n$}
\put(4200,-800){\footnotesize $+16t^6P_n$}
\put(6100,-800){\footnotesize $+28t^6P_n$}
\put(7900,-800){\footnotesize $+4t^6P_n$}
\put(9500,-800){\footnotesize $+2t^4Q_n$}
\put(10800,-800){\footnotesize $+\dots$}
\end{picture}%

\vskip 1.5 cm
\begin{center}
\begin{minipage}{9cm}
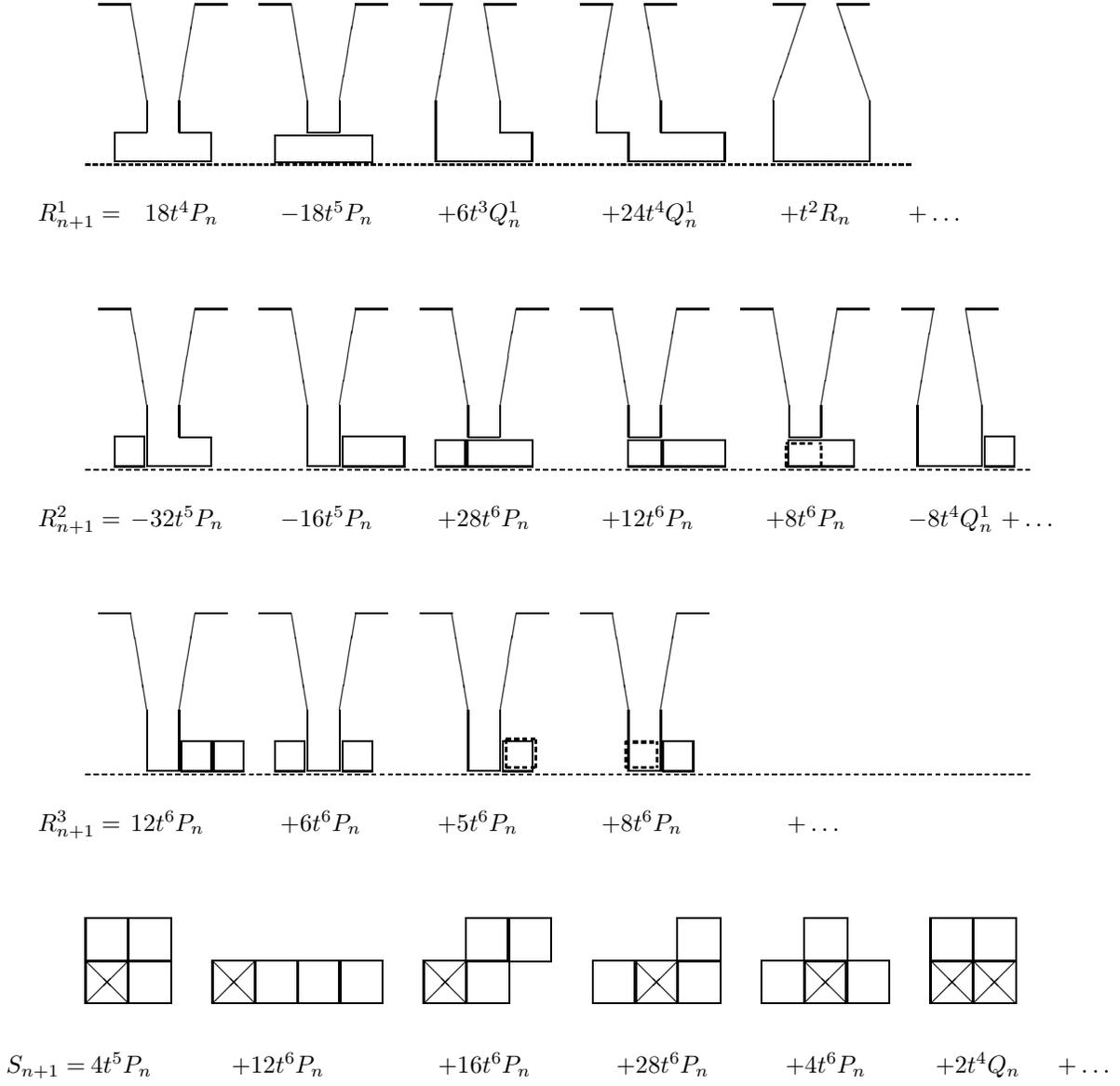
{\footnotesize{\centerline{Fig. 12.  Recursions for $R_{n+1}^1,R_{n+1}^2,R_{n+1}^3$ and $S_{n+1}$.}}} 
\end{minipage}
\end{center}
\vskip .8 cm

Factors larger than the $\pm 2$ in \eqref{m2} and \eqref{m22} are
possible for extensions $\om'$.  At the given orders, though, such factors do not 
appear in our formulas for $\vphi^T(\om')$ or contribute to
the recursion formulas \eqref{induc}, because
the added polymers, $\ga'$, or $\ga'_1$ and $\ga'_2$, do not create new
cycles in the incompatibility graph other than possibly cycles of
length 3, namely $\ga'\not\sim\ga_1\not\sim\ga_2\not\sim\ga'$ or
$\ga'_1\not\sim\ga\not\sim\ga'_2\not\sim\ga'_1$.


\newpage
\section{Diagrams for the 7/8 transition line}\label{78}
\bigskip
\setlength{\unitlength}{1950sp}%
\begin{picture}(12669,2724)(-11,-1873)
\thinlines
{\put(  1,839){\line( 1, 0){450}}
\put(451,839){\line( 0,-1){900}}
\put(451,-61){\line( 1, 0){450}}
\put(901,-61){\line( 0, 1){900}}
\put(901,839){\line( 1, 0){450}}
}%
{\put(1846,839){\line( 1, 0){450}}
\put(2296,839){\line( 0,-1){1305}}
\put(2296,-466){\line( 1, 0){450}}
\put(2746,-466){\line( 0, 1){855}}
\put(2746,389){\line( 1, 0){450}}
\put(3196,389){\line( 0, 1){450}}
\put(3196,839){\line( 1, 0){450}}
}%
{\put(2296,-1411){\line( 0,-1){405}}
\put(2296,-1816){\line( 1, 0){450}}
\put(2746,-1816){\line( 0, 1){405}}
\put(2746,-1411){\line(-1, 0){450}}
}%
{\put(451,-106){\line( 0,-1){405}}
\put(451,-511){\line( 1, 0){450}}
\put(901,-511){\line( 0, 1){405}}
\put(901,-106){\line(-1, 0){450}}
}%
{\put(451,-556){\line( 0,-1){360}}
\put(451,-916){\line( 1, 0){450}}
\put(901,-916){\line( 0, 1){360}}
\put(901,-556){\line(-1, 0){450}}
}%
{\put(451,-961){\line( 0,-1){405}}
\put(451,-1366){\line( 1, 0){450}}
\put(901,-1366){\line( 0, 1){405}}
\put(901,-961){\line(-1, 0){450}}
}%
{\put(6751,-511){\line( 0,-1){855}}
\put(6751,-1366){\line( 1, 0){450}}
\put(7201,-1366){\line( 0, 1){855}}
\put(7201,-511){\line(-1, 0){450}}
}%
{\put(5851,839){\line( 1, 0){450}}
\put(6301,839){\line( 0,-1){450}}
\put(6301,389){\line( 1, 0){450}}
\put(6751,389){\line( 0,-1){855}}
\put(6751,-466){\line( 1, 0){450}}
\put(7201,-466){\line( 0, 1){855}}
\put(7201,389){\line( 1, 0){450}}
\put(7651,389){\line( 0, 1){450}}
\put(7651,839){\line( 1, 0){450}}
}%
{\put(9001,-511){\line( 0,-1){855}}
\put(9001,-1366){\line( 1, 0){450}}
\put(9451,-1366){\line( 0, 1){855}}
\put(9451,-511){\line(-1, 0){450}}
}%
{\put(8551,839){\line( 1, 0){450}}
\put(9001,839){\line( 0,-1){1305}}
\put(9001,-466){\line( 1, 0){450}}
\put(9451,-466){\line( 0, 1){405}}
\put(9451,-61){\line( 1, 0){450}}
\put(9901,-61){\line( 0, 1){900}}
\put(9901,839){\line( 1, 0){450}}
}%
{\put(451,-1411){\line( 0,-1){405}}
\put(451,-1816){\line( 1, 0){450}}
\put(901,-1816){\line( 0, 1){405}}
\put(901,-1411){\line(-1, 0){450}}
}%
{\multiput( 46,-1861)(114,0.00000){111}{\line( 1, 0){ 57.014}}}%
{\put(2296,-961){\line( 0,-1){405}}
\put(2296,-1366){\line( 1, 0){450}}
\put(2746,-1366){\line( 0, 1){405}}
\put(2746,-961){\line(-1, 0){450}}
}%
{\put(2296,-511){\line( 0,-1){405}}
\put(2296,-916){\line( 1, 0){450}}
\put(2746,-916){\line( 0, 1){405}}
\put(2746,-511){\line(-1, 0){450}}
}%
{\put(6751,-1411){\line( 0,-1){405}}
\put(6751,-1816){\line( 1, 0){450}}
\put(7201,-1816){\line( 0, 1){405}}
\put(7201,-1411){\line(-1, 0){450}}
}%
{\put(9001,-1411){\line( 0,-1){405}}
\put(9001,-1816){\line( 1, 0){450}}
\put(9451,-1816){\line( 0, 1){405}}
\put(9451,-1411){\line(-1, 0){450}}
}%
{\put(10801,839){\line( 1, 0){450}}
\put(11251,839){\line( 0,-1){900}}
\put(11251,-61){\line( 1, 0){450}}
\put(11701,-61){\line( 0,-1){900}}
\put(11701,-961){\line(-1, 0){450}}
\put(11251,-961){\line( 0,-1){450}}
\put(11251,-1411){\line( 1, 0){450}}
\put(11701,-1411){\line( 0,-1){405}}
\put(11701,-1816){\line( 0, 1){  0}}
\put(11701,-1816){\line( 1, 0){450}}
\put(12151,-1816){\line( 0, 1){2205}}
\put(12151,389){\line(-1, 0){450}}
\put(11701,389){\line( 0, 1){450}}
\put(11701,839){\line( 1, 0){450}}
}%
{\put(4051,839){\line( 1, 0){450}}
\put(4501,839){\line( 0,-1){450}}
\put(4501,389){\line( 1, 0){450}}
\put(4951,389){\line( 0, 1){450}}
\put(4951,839){\line( 1, 0){450}}
}%
{\put(4501,344){\line( 0,-1){1260}}
\put(4501,-916){\line( 1, 0){450}}
\put(4951,-916){\line( 0, 1){405}}
\put(4951,-511){\line( 1, 0){450}}
\put(5401,-511){\line( 0, 1){450}}
\put(5401,-61){\line(-1, 0){450}}
\put(4951,-61){\line( 0, 1){405}}
\put(4951,344){\line(-1, 0){450}}
\put(4501,344){\line( 0, 1){  0}}
}%
{\put(4501,-961){\line( 0,-1){855}}
\put(4501,-1816){\line( 1, 0){450}}
\put(4951,-1816){\line( 0, 1){855}}
\put(4951,-961){\line(-1, 0){450}}
}%
\put(200,-2400){\footnotesize $\left(n-1\atop4\right)$}
\put(1700,-2400){\footnotesize $-4\left(n-1\atop3\right)$}
\put(3300,-2400){\footnotesize $+8\left(n-1\atop2\right)(n-2)$}
\put(4500,-700){\footnotesize $(1)$}
\put(6100,-2400){\footnotesize $+22\left(n-1\atop2\right)$}
\put(8400,-2400){\footnotesize $+4\left(n-2\atop2\right)$}
\put(10000,-2400){\footnotesize $+64\left(n-2\atop2\right)-16(n-3)$}
\end{picture}%
\vskip 1cm

\begin{picture}(12624,3184)(-11,-1873)
\thinlines
{\put(  1,839){\line( 1, 0){450}}
\put(451,839){\line( 0,-1){1305}}
\put(451,-466){\line( 1, 0){450}}
\put(901,-466){\line( 0, 1){855}}
\put(901,389){\line( 1, 0){450}}
\put(1351,389){\line( 0, 1){450}}
\put(1351,839){\line( 1, 0){450}}
}%
{\put(451,-511){\line( 0,-1){1305}}
\put(451,-1816){\line( 1, 0){450}}
\put(901,-1816){\line( 0, 1){405}}
\put(901,-1411){\line( 1, 0){450}}
\put(1351,-1411){\line( 0, 1){450}}
\put(1351,-961){\line(-1, 0){450}}
\put(901,-961){\line( 0, 1){450}}
\put(901,-511){\line(-1, 0){450}}
}%
{\put(3196,389){\line( 0,-1){855}}
\put(3196,-466){\line( 1, 0){450}}
\put(3646,-466){\line( 0, 1){855}}
\put(3646,389){\line(-1, 0){450}}
}%
{\put(3196,-1816){\framebox(450,1305){}}
}%
{\put(2251,839){\line( 1, 0){450}}
\put(2701,839){\line( 0,-1){900}}
\put(2701,-61){\line( 1, 0){450}}
\put(3151,-61){\line( 0, 1){900}}
\put(3151,839){\line( 1, 0){450}}
}%
{\put(6751,839){\line( 1, 0){450}}
\put(7201,839){\line( 0,-1){900}}
\put(7201,-61){\line( 1, 0){450}}
\put(7651,-61){\line( 0, 1){900}}
\put(7651,839){\line( 1, 0){450}}
}%
{\put(7201,-1816){\framebox(450,1710){}}
}%
\thicklines
{\multiput(7246,-151)(0.00000,-180.00000){3}{\line( 0,-1){ 90.000}}
\multiput(7246,-601)(180.00000,0.00000){3}{\line( 1, 0){ 90.000}}
\multiput(7696,-601)(0.00000,180.00000){3}{\line( 0, 1){ 90.000}}
\multiput(7696,-151)(-180.00000,0.00000){3}{\line(-1, 0){ 90.000}}
}%
\thinlines
{\put(4501,839){\line( 1, 0){450}}
\put(4951,839){\line( 0,-1){900}}
\put(4951,-61){\line( 1, 0){450}}
\put(5401,-61){\line( 0, 1){900}}
\put(5401,839){\line( 1, 0){450}}
}%
{\put(4951,-1816){\framebox(450,1710){}}
}%
{\put(5446,-466){\line( 0,-1){405}}
\put(5446,-871){\line( 1, 0){450}}
\put(5896,-871){\line( 0, 1){405}}
\put(5896,-466){\line(-1, 0){450}}
}%
{\multiput(  1,-1861)(118,0.00000){108}{\line( 1, 0){ 57.349}}
}%
{\put(11251,839){\line( 1, 0){450}}
\put(11701,839){\line( 0,-1){900}}
\put(11701,-61){\line( 1, 0){450}}
\put(12151,-61){\line( 0, 1){900}}
\put(12151,839){\line( 1, 0){450}}
}%
\thicklines
{\multiput(11701,839)(0.00000,128.57143){4}{\line( 0, 1){ 64.286}}
\multiput(11701,1289)(128.57143,0.00000){4}{\line( 1, 0){ 64.286}}
\multiput(12151,1289)(0.00000,-128.57143){4}{\line( 0,-1){ 64.286}}
}%
\thinlines
{\put(11701,-106){\line( 0,-1){855}}
\put(11701,-961){\line( 1, 0){450}}
\put(12151,-961){\line( 0, 1){855}}
\put(12151,-106){\line(-1, 0){450}}
}%
{\put(11701,-1816){\framebox(450,810){}}
}%
{\put(9001,839){\line( 1, 0){450}}
\put(9451,839){\line( 0,-1){1800}}
\put(9451,-961){\line( 1, 0){450}}
\put(9901,-961){\line( 0, 1){450}}
\put(9901,-511){\line( 1, 0){450}}
\put(10351,-511){\line( 0, 1){900}}
\put(10351,389){\line(-1, 0){450}}
\put(9901,389){\line( 0, 1){450}}
\put(9901,839){\line( 1, 0){450}}
}%
{\put(9451,-1816){\framebox(450,810){}}
}%
\put(-200,-2400){\footnotesize $-64\left(n-1\atop2\right)$}
\put(-100,-2800){\footnotesize $+8(n-1)$}
\put(2200,-2400){\footnotesize $+10\left(n-1\atop2\right)$}
\put(2200,-2800){\footnotesize $-2(n-2)$}
\put(4200,-2400){\footnotesize $+10\left(n-1\atop2\right)$}
\put(4200,-2800){\footnotesize $-2(n-2)$}
\put(6650,-2400){\footnotesize $+4(n-2)$}
\put(6650,-500){\footnotesize $(2)$}
\put(9100,-2400){\footnotesize $-16\left(n-2\atop2\right)$}
\put(11400,-2400){\footnotesize $-6\left(n-1\atop2\right)$}
\end{picture}%
\vskip 1.5cm


\setlength{\unitlength}{1750sp}%
\begin{picture}(14424,2724)(-11,-1873)
\thinlines
{\put(  1,839){\line( 1, 0){450}}
\put(451,839){\line( 0,-1){450}}
\put(451,389){\line( 1, 0){450}}
\put(901,389){\line( 0,-1){2205}}
\put(901,-1816){\line( 1, 0){450}}
\put(1351,-1816){\line( 0, 1){855}}
\put(1351,-961){\line( 1, 0){450}}
\put(1801,-961){\line( 0, 1){450}}
\put(1801,-511){\line(-1, 0){450}}
\put(1351,-511){\line( 0, 1){900}}
\put(1351,389){\line( 1, 0){450}}
\put(1801,389){\line( 0, 1){450}}
\put(1801,839){\line( 1, 0){450}}
}%
{\put(2701,839){\line( 1, 0){450}}
\put(3151,839){\line( 0,-1){2250}}
\put(3151,-1411){\line( 1, 0){450}}
\put(3601,-1411){\line( 0,-1){405}}
\put(3601,-1816){\line( 1, 0){450}}
\put(4051,-1816){\line( 0, 1){855}}
\put(4051,-961){\line(-1, 0){450}}
\put(3601,-961){\line( 0, 1){900}}
\put(3601,-61){\line( 1, 0){450}}
\put(4051,-61){\line( 0, 1){900}}
\put(4051,839){\line( 1, 0){450}}
}%
{\put(4951,839){\line( 1, 0){450}}
\put(5401,839){\line( 0,-1){2655}}
\put(5401,-1816){\line( 1, 0){450}}
\put(5851,-1816){\line( 0, 1){405}}
\put(5851,-1411){\line( 1, 0){450}}
\put(6301,-1411){\line( 0, 1){1350}}
\put(6301,-61){\line(-1, 0){450}}
\put(5851,-61){\line( 0, 1){900}}
\put(5851,839){\line( 1, 0){450}}
}%
{\put(8101,-961){\line( 0,-1){855}}
\put(8101,-1816){\line( 1, 0){450}}
\put(8551,-1816){\line( 0, 1){855}}
\put(8551,-961){\line(-1, 0){450}}
}%
{\put(6751,839){\line( 1, 0){450}}
\put(7201,839){\line( 0,-1){450}}
\put(7201,389){\line( 1, 0){900}}
\put(8101,389){\line( 0,-1){1305}}
\put(8101,-916){\line( 1, 0){450}}
\put(8551,-916){\line( 0, 1){1305}}
\put(8551,389){\line( 1, 0){450}}
\put(9001,389){\line( 0, 1){450}}
\put(9001,839){\line( 1, 0){450}}
}%
{\put(10801,-961){\line( 0,-1){855}}
\put(10801,-1816){\line( 1, 0){450}}
\put(11251,-1816){\line( 0, 1){855}}
\put(11251,-961){\line(-1, 0){450}}
}%
{\put(9901,839){\line( 1, 0){450}}
\put(10351,839){\line( 0,-1){450}}
\put(10351,389){\line( 1, 0){450}}
\put(10801,389){\line( 0,-1){1305}}
\put(10801,-916){\line( 1, 0){450}}
\put(11251,-916){\line( 0, 1){855}}
\put(11251,-61){\line( 1, 0){450}}
\put(11701,-61){\line( 0, 1){900}}
\put(11701,839){\line( 1, 0){450}}
}%
{\put(12601,839){\line( 1, 0){450}}
\put(13051,839){\line( 0,-1){2655}}
\put(13051,-1816){\line( 1, 0){450}}
\put(13501,-1816){\line( 0, 1){855}}
\put(13501,-961){\line( 1, 0){450}}
\put(13951,-961){\line( 0, 1){900}}
\put(13951,-61){\line(-1, 0){450}}
\put(13501,-61){\line( 0, 1){450}}
\put(13501,389){\line( 1, 0){450}}
\put(13951,389){\line( 0, 1){450}}
\put(13951,839){\line( 1, 0){450}}
}%
{\multiput(  1,-1861)(114.74104,0.00000){126}{\line( 1, 0){ 57.371}}
}%
\put(300,-2400){\footnotesize $+176n$}
\put(3100,-2400){\footnotesize $+32n$}
\put(5200,-2400){\footnotesize $+8n$}
\put(7450,-2400){\footnotesize $-124n$}
\put(7550,500){\footnotesize $(3)$}
\put(10500,-2400){\footnotesize $-32n$}
\put(13000,-2400){\footnotesize $+32n$}
\end{picture}%
\vskip 1cm

\begin{picture}(14424,3184)(-11,-1873)
\thinlines
{\multiput(  1,-1861)(114.74104,0.00000){126}{\line( 1, 0){ 57.371}}
}%
{\put(  1,839){\line( 1, 0){450}}
\put(451,839){\line( 0,-1){900}}
\put(451,-61){\line(-1, 0){450}}
\put(  1,-61){\line( 0,-1){450}}
\put(  1,-511){\line( 1, 0){450}}
\put(451,-511){\line( 0,-1){1305}}
\put(451,-1816){\line( 1, 0){450}}
\put(901,-1816){\line( 0, 1){1305}}
\put(901,-511){\line( 1, 0){450}}
\put(1351,-511){\line( 0, 1){450}}
\put(1351,-61){\line(-1, 0){450}}
\put(901,-61){\line( 0, 1){900}}
\put(901,839){\line( 1, 0){450}}
}%
{\put(6301,839){\line( 1, 0){450}}
\put(6751,839){\line( 0,-1){450}}
\put(6751,389){\line( 0,-1){1305}}
\put(6751,-916){\line( 1, 0){450}}
\put(7201,-916){\line( 0, 1){405}}
\put(7201,-511){\line( 1, 0){450}}
\put(7651,-511){\line( 0, 1){1350}}
\put(7651,839){\line( 1, 0){450}}
}%
{\put(6751,-961){\line( 0,-1){855}}
\put(6751,-1816){\line( 1, 0){450}}
\put(7201,-1816){\line( 0, 1){855}}
\put(7201,-961){\line(-1, 0){450}}
}%
{\put(1801,839){\line( 1, 0){450}}
\put(2251,839){\line( 0,-1){2655}}
\put(2251,-1816){\line( 1, 0){450}}
\put(2701,-1816){\line( 0, 1){2205}}
\put(2701,389){\line( 1, 0){450}}
\put(3151,389){\line( 0, 1){450}}
\put(3151,839){\line( 1, 0){450}}
}%
{\put(2746,-961){\framebox(405,450){}}
}%
{\put(4051,839){\line( 1, 0){450}}
\put(4501,839){\line( 0,-1){1800}}
\put(4501,-961){\line( 1, 0){450}}
\put(4951,-961){\line( 0, 1){1350}}
\put(4951,389){\line( 1, 0){450}}
\put(5401,389){\line( 0, 1){450}}
\put(5401,839){\line( 1, 0){450}}
}%
{\put(4996,-1816){\framebox(405,1305){}}
}%
{\put(8551,839){\line( 1, 0){450}}
\put(9001,839){\line( 0,-1){900}}
\put(9001,-61){\line( 0,-1){450}}
\put(9001,-511){\line( 0,-1){1305}}
\put(9001,-1816){\line( 1, 0){450}}
\put(9451,-1816){\line( 0, 1){1305}}
\put(9451,-511){\line( 1, 0){450}}
\put(9901,-511){\line( 0, 1){450}}
\put(9901,-61){\line(-1, 0){450}}
\put(9451,-61){\line( 0, 1){900}}
\put(9451,839){\line( 1, 0){450}}
}%
\thicklines
{\multiput(9001,839)(0.00000,128.57143){4}{\line( 0, 1){ 64.286}}
\multiput(9001,1289)(128.57143,0.00000){4}{\line( 1, 0){ 64.286}}
\multiput(9451,1289)(0.00000,-128.57143){4}{\line( 0,-1){ 64.286}}
}%
\thinlines
{\put(10351,839){\line( 1, 0){450}}
\put(10801,839){\line( 0,-1){1800}}
\put(10801,-961){\line( 1, 0){450}}
\put(11251,-961){\line( 0, 1){1350}}
\put(11251,389){\line( 1, 0){450}}
\put(11701,389){\line( 0, 1){450}}
\put(11701,839){\line( 1, 0){450}}
}%
{\put(10801,-1006){\line( 0,-1){810}}
\put(10801,-1816){\line( 1, 0){450}}
\put(11251,-1816){\line( 0, 1){810}}
\put(11251,-1006){\line(-1, 0){450}}
}%
\thicklines
{\multiput(10801,839)(0.00000,128.57143){4}{\line( 0, 1){ 64.286}}
\multiput(10801,1289)(128.57143,0.00000){4}{\line( 1, 0){ 64.286}}
\multiput(11251,1289)(0.00000,-128.57143){4}{\line( 0,-1){ 64.286}}
}%
\thinlines
{\put(13051,-1006){\line( 0,-1){810}}
\put(13051,-1816){\line( 1, 0){450}}
\put(13501,-1816){\line( 0, 1){810}}
\put(13501,-1006){\line(-1, 0){450}}
}%
{\put(12601,839){\line( 1, 0){450}}
\put(13051,839){\line( 0,-1){1800}}
\put(13051,-961){\line( 1, 0){450}}
\put(13501,-961){\line( 0, 1){1800}}
\put(13501,839){\line( 1, 0){900}}
}%
\thicklines
{\multiput(13051,839)(0.00000,128.57143){4}{\line( 0, 1){ 64.286}}
\multiput(13051,1289)(120.00000,0.00000){8}{\line( 1, 0){ 60.000}}
\multiput(13951,1289)(0.00000,-128.57143){4}{\line( 0,-1){ 64.286}}
}%
\put(300,-2400){\footnotesize $+54n$}
\put(2000,-2400){\footnotesize $-20n$}
\put(4400,-2400){\footnotesize $-20n$}
\put(6800,-2400){\footnotesize $-4n$}
\put(8900,-2400){\footnotesize $-48n$}
\put(10900,-2400){\footnotesize $+40n$}
\put(13000,-2400){\footnotesize $+20n$}
\end{picture}%

\vskip 1cm
\begin{center}
\begin{minipage}{9cm}
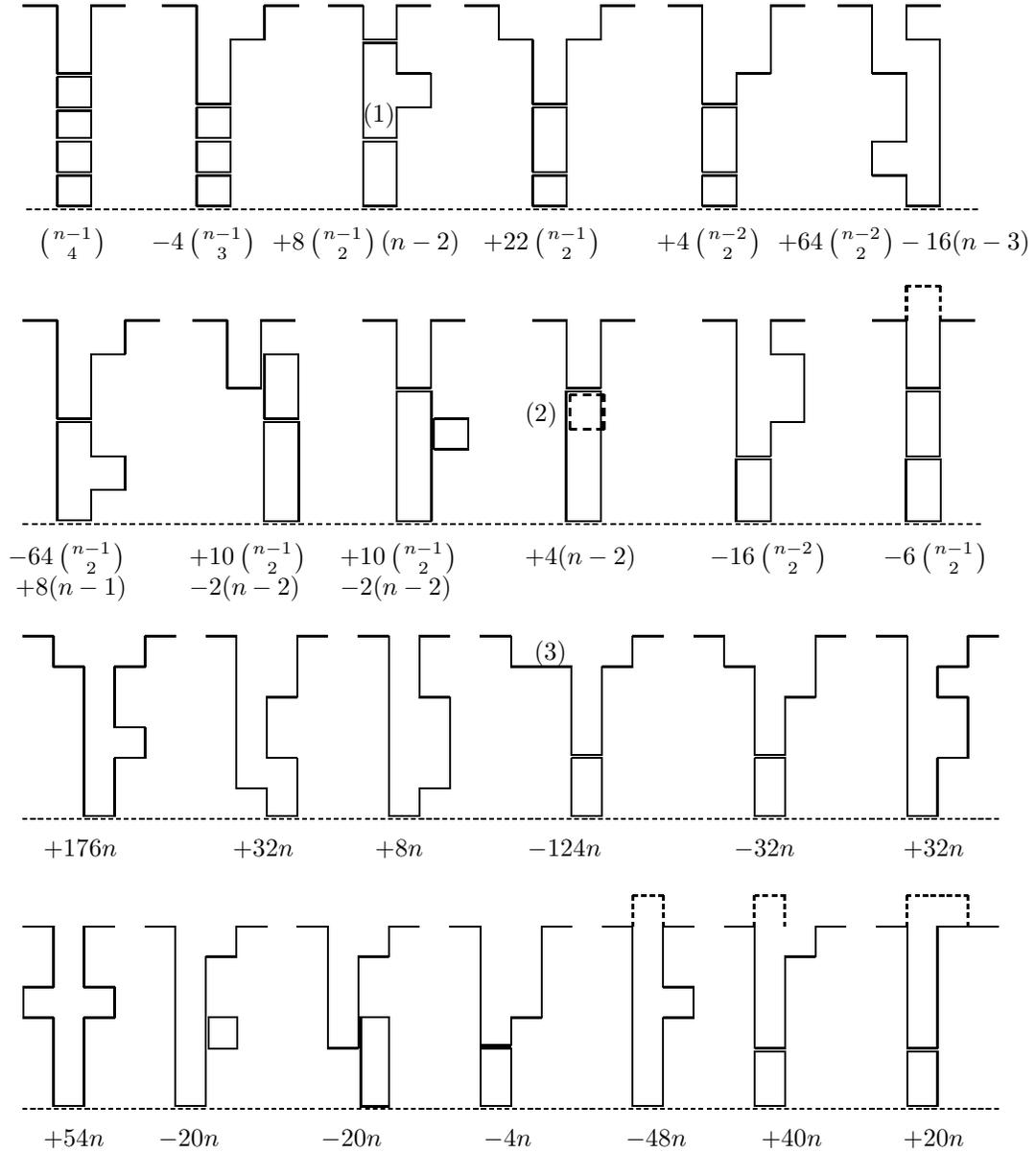
{\footnotesize Fig. 13. $P_n$: $t^4$ terms dependent on $n$. Continuation
downward from levels containing two cubes, as in configuration (1), may be from below
either cube.  Configurations (2) are excluded from the preceding two diagrams.
For (3) see Fig.~14.}
\end{minipage}
\end{center}
\vskip .8 cm 

\begin{picture}(11724,1846)(889,-995)
\thinlines
{\put(901,-961){\framebox(450,1800){}}
}%
{\put(901,-511){\line( 1, 1){450}}
}%
{\put(901,-61){\line( 1,-1){450}}
}%
{\put(2701,389){\line( 0,-1){1350}}
\put(2701,-961){\line( 1, 0){900}}
\put(3601,-961){\line( 0, 1){450}}
\put(3601,-511){\line(-1, 0){450}}
\put(3151,-511){\line( 0, 1){900}}
\put(3151,389){\line(-1, 0){450}}
\put(2701,389){\line( 0, 1){  0}}
}%
{\put(4951,389){\line( 0,-1){900}}
\put(4951,-511){\line( 1, 0){450}}
\put(5401,-511){\line( 0,-1){450}}
\put(5401,-961){\line( 1, 0){450}}
\put(5851,-961){\line( 0, 1){900}}
\put(5851,-61){\line(-1, 0){450}}
\put(5401,-61){\line( 0, 1){450}}
\put(5401,389){\line(-1, 0){450}}
\put(4951,389){\line( 0, 1){  0}}
}%
{\put(7201,389){\line( 0,-1){1350}}
\put(7201,-961){\line( 1, 0){450}}
\put(7651,-961){\line( 0, 1){450}}
\put(7651,-511){\line( 1, 0){450}}
\put(8101,-511){\line( 0, 1){450}}
\put(8101,-61){\line(-1, 0){450}}
\put(7651,-61){\line( 0, 1){450}}
\put(7651,389){\line(-1, 0){450}}
\put(7201,389){\line( 0, 1){  0}}
}%
{\put(9451,389){\line( 0,-1){1350}}
\put(9451,-961){\line( 1, 0){900}}
\put(10351,-961){\line( 0, 1){900}}
\put(10351,-61){\line(-1, 0){450}}
\put(9901,-61){\line( 0, 1){450}}
\put(9901,389){\line(-1, 0){450}}
\put(9451,389){\line( 0, 1){  0}}
}%
{\put(11701,-961){\framebox(900,1350){}}
}%
{\put(2701,-61){\line( 1,-1){450}}
}%
{\put(5423,-83){\line( 1,-1){450}}
}%
{\put(7673,-83){\line( 1,-1){450}}
}%
{\put(9473,-83){\line( 1,-1){450}}
}%
{\put(11723,-533){\line( 1,-1){450}}
}%
{\put(2723,-488){\line( 1, 1){450}}
}%
{\put(5401,-511){\line( 1, 1){450}}
}%
{\put(7651,-511){\line( 1, 1){450}}
}%
{\put(9451,-511){\line( 1, 1){450}}
}%
{\put(11701,-961){\line( 1, 1){450}}
}%
{\put(901,-61){\line( 1, 0){450}}
}%
{\put(901,-511){\line( 1, 0){450}}
}%
{\put(2701,-61){\line( 1, 0){450}}
}%
{\put(2701,-511){\line( 1, 0){450}}
}%
{\put(4951,-61){\line( 1, 0){450}}
}%
{\put(5401,-511){\line( 1, 0){450}}
}%
{\put(7201,-61){\line( 1, 0){450}}
}%
{\put(7201,-511){\line( 1, 0){450}}
}%
{\put(9451,-61){\line( 1, 0){450}}
}%
{\put(9451,-511){\line( 1, 0){900}}
}%
{\put(3151,-511){\line( 0,-1){450}}
}%
{\put(5401,-61){\line( 0,-1){450}}
}%
{\put(7651,-16){\line( 0,-1){450}}
}%
{\put(11701,-61){\line( 1, 0){900}}
}%
{\put(11701,-511){\line( 1, 0){900}}
}%
{\put(12151,389){\line( 0,-1){1350}}
}%
{\put(9901,-61){\line( 0,-1){900}}
}%
{\put(901,389){\line( 1, 0){450}}
}%
\put(1100,-1600){\footnotesize $8$}
\put(2500,-1600){\footnotesize $+\ 32$}
\put(4900,-1600){\footnotesize $+\  16$}
\put(7300,-1600){\footnotesize $+\  16$}
\put(9400,-1600){\footnotesize $+\  40$}
\put(11800,-1600){\footnotesize $+\  12$}
\put(13500,-1600){\footnotesize $=\  124$}
\end{picture}%

\vskip 1cm
\begin{center}
\begin{minipage}{9cm}
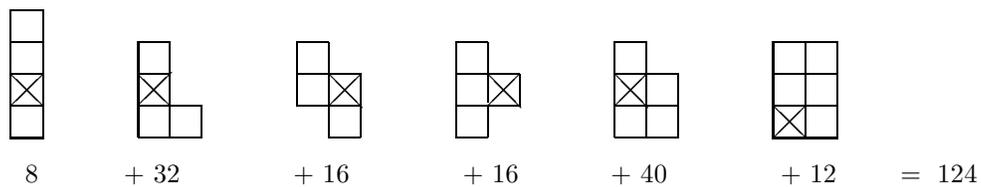
{\footnotesize Fig. 14. Terms contributing to (3) in Fig. 13.  Top view; $\times$ represents a possible location of the column below.}
\end{minipage}
\end{center}
\vskip .8 cm 

\begin{picture}(12624,2724)(34,-1873)
\thinlines
{\put(451,-106){\line( 0,-1){405}}
\put(451,-511){\line( 1, 0){450}}
\put(901,-511){\line( 0, 1){405}}
\put(901,-106){\line(-1, 0){450}}
}%
{\put(451,-556){\line( 0,-1){360}}
\put(451,-916){\line( 1, 0){450}}
\put(901,-916){\line( 0, 1){360}}
\put(901,-556){\line(-1, 0){450}}
}%
{\put(451,-961){\line( 0,-1){405}}
\put(451,-1366){\line( 1, 0){450}}
\put(901,-1366){\line( 0, 1){405}}
\put(901,-961){\line(-1, 0){450}}
}%
{\put(451,-1411){\line( 0,-1){405}}
\put(451,-1816){\line( 1, 0){450}}
\put(901,-1816){\line( 0, 1){405}}
\put(901,-1411){\line(-1, 0){450}}
}%
{\multiput( 46,-1861)(114.02715,0.00000){111}{\line( 1, 0){ 57.014}}
}%
{\put(451,839){\line( 0,-1){900}}
\put(451,-61){\line( 1, 0){450}}
\put(901,-61){\line( 0, 1){900}}
\put(901,839){\line(-1, 0){450}}
}%
{\put(2701,839){\line( 0,-1){450}}
\put(2701,389){\line( 1, 0){450}}
\put(3151,389){\line( 0, 1){450}}
\put(3151,839){\line(-1, 0){450}}
}%
{\put(2701,344){\line( 0,-1){1260}}
\put(2701,-916){\line( 1, 0){450}}
\put(3151,-916){\line( 0, 1){405}}
\put(3151,-511){\line( 1, 0){450}}
\put(3601,-511){\line( 0, 1){450}}
\put(3601,-61){\line(-1, 0){450}}
\put(3151,-61){\line( 0, 1){405}}
\put(3151,344){\line(-1, 0){450}}
}%
{\put(2701,-961){\line( 0,-1){855}}
\put(2701,-1816){\line( 1, 0){450}}
\put(3151,-1816){\line( 0, 1){855}}
\put(3151,-961){\line(-1, 0){450}}
}%
{\put(5401,839){\line( 0,-1){1305}}
\put(5401,-466){\line( 1, 0){450}}
\put(5851,-466){\line( 0, 1){855}}
\put(5851,389){\line( 1, 0){450}}
\put(6301,389){\line( 0, 1){450}}
\put(6301,839){\line(-1, 0){900}}
}%
{\put(5401,-511){\line( 0,-1){855}}
\put(5401,-1366){\line( 1, 0){450}}
\put(5851,-1366){\line( 0, 1){855}}
\put(5851,-511){\line(-1, 0){450}}
}%
{\put(5401,-1411){\line( 0,-1){405}}
\put(5401,-1816){\line( 1, 0){450}}
\put(5851,-1816){\line( 0, 1){405}}
\put(5851,-1411){\line(-1, 0){450}}
}%
{\put(10801,839){\line( 0,-1){450}}
\put(10801,389){\line( 1, 0){450}}
\put(11251,389){\line( 0,-1){1350}}
\put(11251,-961){\line(-1, 0){450}}
\put(10801,-961){\line( 0,-1){450}}
\put(10801,-1411){\line( 1, 0){450}}
\put(11251,-1411){\line( 0,-1){405}}
\put(11251,-1816){\line( 1, 0){450}}
\put(11701,-1816){\line( 0, 1){2655}}
\put(11701,839){\line(-1, 0){900}}
}%
{\put(8101,839){\line( 0,-1){900}}
\put(8101,-61){\line( 1, 0){450}}
\put(8551,-61){\line( 0,-1){900}}
\put(8551,-961){\line(-1, 0){450}}
\put(8101,-961){\line( 0,-1){450}}
\put(8101,-1411){\line( 1, 0){450}}
\put(8551,-1411){\line( 0,-1){405}}
\put(8551,-1816){\line( 1, 0){450}}
\put(9001,-1816){\line( 0, 1){2205}}
\put(9001,389){\line(-1, 0){450}}
\put(8551,389){\line( 0, 1){450}}
\put(8551,839){\line(-1, 0){450}}
}%
\put(200,-2400){\footnotesize $\left(n-1\atop4\right)$}
\put(1600,-2400){\footnotesize $+8\left(n-1\atop2\right)(n-2)$}
\put(4700,-2400){\footnotesize $+4\left(n-1\atop2\right)$}
\put(6500,-2400){\footnotesize $+64\left(n-2\atop2\right)-16(n-3)$}
\put(10400,-2400){\footnotesize $+32(n-2)+8$}
{\multiput(  1,899)(114.69767,0.00000){108}{\line( 1, 0){ 57.349}}}%
\end{picture}%

\vskip 1.3cm


\begin{picture}(12354,2724)(-11,-1873)
\thinlines
{\multiput(  1,-1861)(114.69767,0.00000){108}{\line( 1, 0){ 57.349}}
}%
{\put(451,-1816){\framebox(450,810){}}
}%
{\put(451,839){\line( 0,-1){1800}}
\put(451,-961){\line( 1, 0){450}}
\put(901,-961){\line( 0, 1){450}}
\put(901,-511){\line( 1, 0){450}}
\put(1351,-511){\line( 0, 1){900}}
\put(1351,389){\line(-1, 0){450}}
\put(901,389){\line( 0, 1){450}}
\put(901,839){\line(-1, 0){450}}
}%
{\put(6301,-1816){\framebox(450,1305){}}
}%
{\put(6301,389){\line( 0,-1){855}}
\put(6301,-466){\line( 1, 0){450}}
\put(6751,-466){\line( 0, 1){855}}
\put(6751,389){\line(-1, 0){450}}
}%
{\put(5806,839){\line( 0,-1){900}}
\put(5806,-61){\line( 1, 0){450}}
\put(6256,-61){\line( 0, 1){900}}
\put(6256,839){\line(-1, 0){450}}
}%
{\put(3151,839){\line( 0,-1){1800}}
\put(3151,-961){\line( 1, 0){450}}
\put(3601,-961){\line( 0, 1){900}}
\put(3601,-61){\line( 1, 0){450}}
\put(4051,-61){\line( 0, 1){900}}
\put(4051,839){\line(-1, 0){900}}
}%
{\put(3151,-1816){\framebox(450,810){}}
}%
{\put(11251,-1816){\framebox(450,1710){}}
}%
{\put(11251,839){\line( 0,-1){900}}
\put(11251,-61){\line( 1, 0){450}}
\put(11701,-61){\line( 0, 1){900}}
\put(11701,839){\line(-1, 0){450}}
}%
\thicklines
{\multiput(11296,-151)(0.00000,-180.00000){3}{\line( 0,-1){ 90.000}}
\multiput(11296,-601)(180.00000,0.00000){3}{\line( 1, 0){ 90.000}}
\multiput(11746,-601)(0.00000,180.00000){3}{\line( 0, 1){ 90.000}}
\multiput(11746,-151)(-180.00000,0.00000){3}{\line(-1, 0){ 90.000}}
}%
\thinlines
{\put(9001,839){\line( 0,-1){900}}
\put(9001,-61){\line( 1, 0){450}}
\put(9451,-61){\line( 0, 1){900}}
\put(9451,839){\line(-1, 0){450}}
}%
{\put(9001,-1816){\framebox(450,1710){}}
}%
{\put(9496,-511){\line( 0,-1){405}}
\put(9496,-916){\line( 1, 0){450}}
\put(9946,-916){\line( 0, 1){405}}
\put(9946,-511){\line(-1, 0){450}}
}%
\put(00,-2400){\footnotesize $-16\left(n-2\atop2\right)$}
\put(2650,-2400){\footnotesize $-4(n-2)$}
\put(5700,-2400){\footnotesize $+10\left(n-1\atop2\right)$}
\put(5700,-2800){\footnotesize $-2(n-2)$}
\put(8400,-2400){\footnotesize $+10\left(n-1\atop2\right)$}
\put(8400,-2800){\footnotesize $-2(n-2)$}
\put(10650,-2400){\footnotesize $+4(n-2)$}
{\multiput(  1,899)(114.69767,0.00000){108}{\line( 1, 0){ 57.349}}}%
\end{picture}%

\vskip 1.5cm

\begin{picture}(12354,3174)(-11,-1873)
\thinlines
{\multiput(  1,-1861)(114.69767,0.00000){108}{\line( 1, 0){ 57.349}}
}%
{\put(451,794){\line( 0,-1){2610}}
\put(451,-1816){\line( 1, 0){450}}
\put(901,-1816){\line( 0, 1){855}}
\put(901,-961){\line( 1, 0){450}}
\put(1351,-961){\line( 0, 1){1350}}
\put(1351,389){\line(-1, 0){450}}
\put(901,389){\line( 0, 1){405}}
\put(901,794){\line(-1, 0){450}}
}%
{\multiput(  1,839)(114.69767,0.00000){108}{\line( 1, 0){ 57.349}}}%
{\put(11251,1289){\line( 0,-1){1800}}
\put(11251,-511){\line(-1, 0){450}}
\put(10801,-511){\line( 0,-1){1305}}
\put(10801,-1816){\line( 1, 0){450}}
\put(11251,-1816){\line( 0, 1){855}}
\put(11251,-961){\line( 1, 0){450}}
\put(11701,-961){\line( 0, 1){2250}}
\put(11701,1289){\line(-1, 0){450}}
}%
{\put(8551,-1816){\framebox(450,1305){}}
}%
{\put(8551,389){\line( 0,-1){855}}
\put(8551,-466){\line( 1, 0){450}}
\put(9001,-466){\line( 0, 1){855}}
\put(9001,389){\line(-1, 0){450}}
}%
{\put(8551,1289){\line( 0,-1){855}}
\put(8551,434){\line( 1, 0){450}}
\put(9001,434){\line( 0, 1){855}}
\put(9001,1289){\line(-1, 0){450}}
}%
{\put(3646,794){\line( 0,-1){405}}
\put(3646,389){\line( 1, 0){450}}
\put(4096,389){\line( 0, 1){405}}
\put(4096,794){\line(-1, 0){450}}
}%
{\put(3151,794){\line( 0,-1){855}}
\put(3151,-61){\line( 1, 0){450}}
\put(3601,-61){\line( 0, 1){855}}
\put(3601,794){\line(-1, 0){450}}
}%
{\put(3151,-1816){\framebox(450,1710){}}
}%
{\put(5851,-1816){\framebox(450,2160){}}
}%
{\put(5851,794){\line( 0,-1){405}}
\put(5851,389){\line( 1, 0){450}}
\put(6301,389){\line( 0, 1){405}}
\put(6301,794){\line(-1, 0){450}}
}%
\thicklines
{\multiput(5941,704)(0.00000,-180.00000){3}{\line( 0,-1){ 90.000}}
\multiput(5941,254)(180.00000,0.00000){3}{\line( 1, 0){ 90.000}}
\multiput(6391,254)(0.00000,180.00000){3}{\line( 0, 1){ 90.000}}
\multiput(6391,704)(-180.00000,0.00000){3}{\line(-1, 0){ 90.000}}
}%
\put(00,-2400){\footnotesize $+8(n-4)$}
\put(2150,-2400){\footnotesize $+5(n-1)-1$}
\put(5700,-2400){\footnotesize $+1$}
\put(8200,-2400){\footnotesize $-\left(n-1\atop2\right)$}
\put(10450,-2400){\footnotesize $-8(n-2)$}
\end{picture}%

\vskip 1cm
\begin{center}
\begin{minipage}{9cm}
{\footnotesize Fig. 15. $\ti P_n$: $t^5$ terms dependent on $n$.}
\end{minipage}
\end{center}
\vskip .8 cm 
\newpage
{\bf Acknowledgments. }
The possibility to meet offered to the authors by the 
Laboratoire de Physique Th{\'e}orique et Modelisation, 
Universit{\'e} de Cergy-Pontoise, and 
the Centre de Physique Th{\'e}orique, CNRS, Marseille, 
is gratefully acknowledged.  The research of K.\ A. was supported by NSF grants
DMS-0405915 and DMS-0804934.


\end{document}